\documentclass[aps,nofootinbib,notitlepage,longbibliography]{revtex4-1}

\usepackage{amsmath}
\usepackage{amsfonts}
\usepackage{amssymb}
\usepackage{mathrsfs}
\usepackage{mathtools}
\usepackage{multirow}

\usepackage{hyperref}

\usepackage[font=small]{caption}
\usepackage[section]{placeins}
\usepackage{accents}
\usepackage{color}

\DeclareSymbolFont{bbold}{U}{bbold}{m}{n}
\DeclareSymbolFontAlphabet{\mathbbold}{bbold}

\newcommand{\ud}{\mathrm{d}}

\newcommand{\pb}[1]{\,\mbox{}_{#1}}
\newcommand{\pp}[1]{\,\mbox{}^{#1}}
\newcommand{\scri}{\mathscr{I}}
\newcommand{\lie}{\pounds}
\newcommand{\var}{\mathchar'26\mkern-9mu \delta}
\newcommand{\dual}{{}^*}
\newcommand{\npdelta}{\delta}
\newcommand{\npDelta}{\mathbbold{\Delta}}

\renewcommand{\Im}{\operatorname{Im}}

\allowdisplaybreaks[3]

\numberwithin{equation}{section}

\begin{document}

\title{Conserved currents for electromagnetic fields in the Kerr spacetime}
\author{Alexander M. Grant, \'{E}anna \'{E}. Flanagan}
\affiliation{Department of Physics, Cornell University, Ithaca, NY 14853, USA}

\begin{abstract}
  We construct a variety of conserved currents for test electromagnetic fields on a Kerr background.
  Our procedure, which involves the symplectic product for electromagnetism and symmetry operators, generates the conserved currents given by Andersson, B\"ackdahl, and Blue (2015), as well as a new conserved current.
  These currents reduce to the sum of (positive powers of) the Carter constants of photons in the geometric optics limit, and generalize the current for scalar fields discovered by Carter (1977) involving the Killing tensor.
  We furthermore show that the fluxes of our new current through null infinity and the horizon are finite.
\end{abstract}

\maketitle

\tableofcontents

\newpage

\section{Introduction}

Freely falling point particles in the Kerr spacetime possess a constant of motion, the \emph{Carter constant} $K$, associated with the existence of a Killing tensor $K_{ab}$ in this spacetime~\cite{PhysRev.174.1559, 1970CMaPh..18..265W}:

\begin{equation} \label{eqn:carter}
  K \equiv K_{ab} p^a p^b,
\end{equation}
where

\begin{equation}
  \nabla_{(a} K_{bc)} = 0.
\end{equation}
Moreover, there exist generalizations of this Carter constant for charged particles in the Kerr-Newman spacetime~\cite{PhysRev.174.1559}, as well as for spinning particles, to linear order in spin~\cite{1983RSPSA.385..229R}.
This constant of motion, along with the energy $E$ and axial angular momentum $L_z$, allows for the solution of the geodesic equations in terms of first integrals.

Unlike the Carter constant, the energy and axial angular momentum of the point particle are associated with Killing vectors $t^a$ and $\phi^a$, and so can be related to conserved currents $T^a{}_b t^b$ and $T^a{}_b \phi^b$, for any field theory with a stress-energy tensor $T_{ab}$.
Moreover, the fluxes of these conserved currents determine the evolution of the energy and axial angular momentum, respectively, of a point particle that couples to this field theory.
Is there any similar story for the Carter constant~\eqref{eqn:carter}?

In~\cite{Grant:2015xqa}, we showed that one cannot construct conserved currents that are related to the Carter constant in this way from arbitrary stress-energy tensors.
In particular, we showed that there is no functional of a generic stress-energy tensor and its derivatives on a spacelike hypersurface $\Sigma$ that reduces to the Carter constant when the stress-energy tensor describes a point particle, and is independent of $\Sigma$ whenever the stress-energy tensor is conserved.\footnote{Note that, using the Killing-Yano tensor $f_{ab}$ (see the discussion below in section~\ref{sec:teukolsky}),~\cite{Witzany2017} has constructed a conserved current for generic theories using the stress-energy tensor.
  However, the existence of such a current does not violate the result in~\cite{Grant:2015xqa}, since it does not reduce to the Carter constant for a point particle.}

This result does not eliminate the existence of conserved currents related to the Carter constant that are not constructed from a stress-energy tensor.
In fact, it is known that scalar fields in the Kerr spacetime possess a conserved current that generalizes the Carter constant~\cite{PhysRevD.16.3395}, in the following sense: in the geometric optics limit, the integral of this current over a surface is given by the the sum of the Carter constants of all of the scalar quanta that pass through this surface.
A similar current also exists for spin-$1/2$ fields~\cite{PhysRevD.19.1093}.
One may ask if a current of this sort exists for other field theories.

The main result of this paper is the construction of conserved currents for electromagnetic fields in the Kerr spacetime that are associated with the Carter constant in the geometric optics limit.
Our method of constructing currents, using symmetry operators which map the space of solutions to itself, together with bilinear currents, is the same as that of~\cite{PhysRevD.16.3395} and~\cite{PhysRevD.19.1093}.
However, we focus primarily on symmetry operators and currents which readily generalize to linearized gravity on the Kerr background, and in particular the symmetry operators which we consider are not obvious generalizations of those in~\cite{PhysRevD.16.3395} and~\cite{PhysRevD.19.1093}.

We consider two currents, which we define in Section~\ref{sec:new_currents}: we denote them by $\pb{\mathcal A} j^a [\var \boldsymbol{A}]$ and $\pb{\pm 1} j^a [\var \boldsymbol{A}]$, defined in equations~\eqref{eqn:A_j} and~\eqref{eqn:pm_1_j}, respectively.
Here $\var \boldsymbol{A}$ is a perturbation to the electromagnetic vector potential.
The latter of these currents is new, whereas the first was previous defined in~\cite{Andersson:2015xla}.
For $\pb{\mathcal A} j^a [\var \boldsymbol{A}]$, we find that the geometric optics limits of this current is proportional to $K$ [see equation~\eqref{eqn:A_j_optics}], whereas for $\pb{\pm 1} j^a [\var \boldsymbol{A}]$ the limit is proportional to $K^2$ [see equation~\eqref{eqn:pm_1_j_optics}].

In addition to computing the geometric optics limit of these currents, we also consider the fluxes of these currents through null infinity and the horizon.
This is motivated by the idea that these fluxes might allow one to determine the evolution of the Carter constant of a charged particle in Kerr that is emitting radiation.
We show that the fluxes of $\pb{\pm 1} j^a [\var \boldsymbol{A}]$ are finite, giving explicit expressions for the fluxes in equations~\eqref{eqn:fluxes_pm_1_Q}.
The flux of $\pb{\mathcal A} j^a [\var \boldsymbol{A}]$ is infinite at null infinity, so it would not be useful for determining the evolution of the Carter constant.

The layout of this paper is as follows: in Section~\ref{sec:review}, we review the formalism of electromagnetic perturbations in the Kerr spacetime, covering the spinor, Newman-Penrose, and Teukolsky formalisms.
We then, in Section~\ref{sec:symmetry_operators}, use these formalisms to construct symmetry operators which map the space of solutions to Maxwell's equations to itself.
Section~\ref{sec:currents} reviews the currents that were constructed in~\cite{Andersson:2015xla}, along with the symplectic product construction for an arbitrary Lagrangian field theory (see~\cite{Burnett57}, for example).
We then consider the symplectic product for electromagnetism, and show the circumstances under which the symplectic product procedure reduces to the procedure for constructing currents given in~\cite{Andersson:2015xla}.
In Section~\ref{sec:geometric_optics} we review the geometric optics limit for electromagnetism, and derive the geometric optics limits of the currents we have defined.
In Section~\ref{sec:fluxes}, we compute the fluxes for these currents at null infinity and the horizon.

In this paper we use the following conventions: we generally follow the conventions of Penrose and Rindler~\cite{penrose1987spinors, penrose1988spinors}, in particular the $(+, -, -, -)$ metric signature convention and their convention for the sign of the Riemann tensor.
Tensors with indices removed are in a bold face font, as is typically done with differential forms.
For a linear (differential) operator $\mathcal{T}_{a_1 \cdots a_p}{}^{b_1 \cdots b_q}$ that maps tensors of rank $q$ to rank $p$, we denote $\mathcal{T}_{a_1 \cdots a_p}{}^{b_1 \cdots b_q} S_{b_1 \cdots b_q}$ by $\boldsymbol{\mathcal T} \cdot \boldsymbol{S}$ when the indices are removed.
Moreover, we use the convention where linear differential operators are applied in a right associative manner, that is,

\begin{equation}
  \boldsymbol{\mathcal T} \cdot \boldsymbol{\mathcal S} \cdot \boldsymbol{\mathcal R} = \boldsymbol{\mathcal T} \cdot (\boldsymbol{\mathcal S} \cdot \boldsymbol{\mathcal R}).
\end{equation}
Finally, in all calculations we implicitly use the soldering forms $\sigma_a{}^{AA'}$ which form the isomorphism between the tangent space and the space of Hermitian spinors~\cite{penrose1987spinors}.
That is, we implicitly associate indices $a$ with $AA'$, $b$ with $BB'$, etc. on two sides of an equation.

\section{Electromagnetic perturbations on a Kerr background} \label{sec:review}

An electromagnetic perturbation on a fixed Kerr background is given by a tensor $\var F_{ab}$ satisfying

\begin{equation} \label{eqn:maxwell}
  \nabla^a \var F_{ab} = 4\pi \var J_b,
\end{equation}
where $\var J_a$ is a linearized source current.
The ``$\var$'' represents the fact that this perturbation is obtained from a one-parameter family of solutions $[F_{ab} (\lambda), g_{ab} (\lambda)]$ to the full Einstein-Maxwell equations, defining the variation $\var \boldsymbol{Q} (\lambda)$ of a quantity $\boldsymbol{Q}$ by\footnote{Note that we are using a non-standard symbol, $\var$, to denote variations instead of $\delta$.
  This is due to an unfortunate clash of notation with the Newman-Penrose directional derivatives, which also include an operator denoted with $\delta$.}

\begin{equation}
  \var \boldsymbol{Q} \equiv \left.\frac{\ud \boldsymbol{Q}}{\ud \lambda}\right|_{\lambda = 0}.
\end{equation}
We are allowed to consider solely electromagnetic perturbations because $\var T_{ab} = 0$, and so we can consistently set $\var g_{ab} = 0$.
Considering electromagnetic perturbations as variations will be useful in Section~\ref{sec:symplectic} below.

The homogeneous Maxwell's equation is given by $\ud \var \boldsymbol{F} = 0$, and so there a 1-form potential $\var \boldsymbol{A}$ such that $\var \boldsymbol{F} = \ud \var \boldsymbol{A}$.
We denote the operator that maps vector potentials into the corresponding sources by $\pb{1} \boldsymbol{\mathcal{E}}$:

\begin{equation} \label{eqn:maxwell_operator_def}
  \pb{1} \mathcal{E}_a{}^b \equiv 2(\delta_a{}^b \nabla^c \nabla_c - \nabla^b \nabla_a),
\end{equation}
such that, from equation~\eqref{eqn:maxwell},

\begin{equation} \label{eqn:maxwell_operator}
  \pb{1} \mathcal{E}_a{}^b \var A_b = 8\pi \var J_a.
\end{equation}

Since the theory of electromagnetic perturbations is linear, we will consider the complexified solution space for convenience, in which case the Faraday tensor has six complex components.
In this case, $\var J_a$ will also be complex.

\subsection{Spinor and Newman-Penrose formalism}

In order to discuss the spinor and Newman-Penrose formalism of electromagnetic perturbations on a Kerr background, we use the conventions and terminology of~\cite{penrose1987spinors, penrose1988spinors}.
By the symmetries of the Faraday tensor, $\var F_{ab}$ can be decomposed into symmetric spinor fields $\var \phi_{AB}$ and $\var \bar{\chi}_{A'B'}$ via

\begin{equation}
  \var F_{ab} \equiv \epsilon_{A'B'} \var \phi_{AB} + \epsilon_{AB} \var \bar{\chi}_{A'B'} \equiv \pp{-} [\var F_{ab}] + \pp{+} [\var F_{ab}].
\end{equation}
The operations $\pp{+} [\cdot]$ and $\pp{-} [\cdot]$ take the \emph{self-dual} and \emph{anti-self-dual} parts of $\var F_{ab}$, respectively, since taking the Hodge dual of these two parts yields

\begin{equation} \label{eqn:(anti)self_dual}
  \dual \pp{\pm} [\var \boldsymbol{F}] = \pm i \pp{\pm} [\var \boldsymbol{F}].
\end{equation}

In terms of spinors, the relationship between the vector potential and the Faraday tensor is given by

\begin{subequations} \label{eqn:faraday_spinor}
  \begin{align}
    \var \phi_{AB} &= \nabla_{(B}{}^{A'} \var A_{A)A'} \equiv \pb{1} \mathcal{F}_{AB}{}^c \var A_c, \\
    \var \bar{\chi}_{A'B'} &= \nabla_{(B'}{}^{A} \var A_{A')A} \equiv \overline{\pb{1} \mathcal{F}_{A'B'}{}^c} \var A_c,
  \end{align}
\end{subequations}
where the second equalities define the operator $\pb{1} \mathcal{F}_{AB}{}^c$, which we will need frequently below.
When we are considering real solutions, $\var \phi_{AB} = \var \chi_{AB}$.
The inhomogeneous Maxwell's equation becomes, in spinor language,

\begin{equation} \label{eqn:maxwell_spinor}
  \nabla^{AB'} \var \phi_A{}^B = 2 \pi \var J^{BB'}, \qquad \nabla^{A'B} \var \bar{\chi}_{A'}{}^{B'} = 2 \pi \var \bar{J}^{BB'}.
\end{equation}

The Newman-Penrose approach is to introduce a spin basis $(o, \iota)$, that is, a pair of spinors $o^A$ and $\iota^A$ such that $o_A \iota^A = -o^A \iota_A = 1$.
The two spinors $\phi_{AB}$ and $\chi_{AB}$ can be decomposed along this spin basis into six complex scalars via the following procedure: we define, for any symmetric spinor $S_{AB}$,

\begin{equation} \label{eqn:scalar_def}
  S_0 \equiv S_{AB} o^A o^B, \quad S_1 \equiv S_{AB} \iota^A o^B, \quad S_2 \equiv S_{AB} \iota^A \iota^B
\end{equation}
The six complex scalars are then $\phi_0$, $\phi_1$, $\phi_2$, $\bar{\chi}_0$, $\bar{\chi}_1$, and $\bar{\chi}_2$.
One can also consider the null tetrad given by

\begin{equation} \label{eqn:np_tetrad}
  l^a = o^A \bar{o}^{A'}, \quad n^a = \iota^A \bar{\iota}^{A'}, \quad m^a = o^A \bar{\iota}^{A'}.
\end{equation}
These definitions, along with $o_A \iota^A = 1$, imply that $l_a n^a = -m_a \bar{m}^a = 1$, with all other contractions between $\ell^a$, $n^a$, $m^a$, and $\bar{m}^a$ being zero.
Denote directional derivatives along the members of this tetrad by $D \equiv l^a \nabla_a$, $\npDelta \equiv n^a \nabla_a$, $\delta \equiv m^a \nabla_a$, and $\bar{\delta} \equiv \bar{m}^a \nabla_a$, and define the twelve spin coefficients

\begin{equation}
    \begin{aligned}
    D o_A &= \epsilon o_A - \kappa \iota_A, \qquad &D \iota_A &= \pi o_A - \epsilon \iota_A, \\
    \npDelta o_A &= \gamma o_A - \tau \iota_A, \qquad &\npDelta \iota_A &= \nu o_A - \gamma \iota_A, \\
    \npdelta o_A &= \beta o_A - \sigma \iota_A, \qquad &\npdelta \iota_A &= \mu o_A - \beta \iota_A, \\
    \bar{\npdelta} o_A &= \alpha o_A - \rho \iota_A, \qquad &\bar{\npdelta} \iota_A &= \lambda o_A - \alpha \iota_A.
  \end{aligned}
\end{equation}
Using the tetrad~\eqref{eqn:np_tetrad}, the scalars $\var \phi_i$ and $\var \bar{\chi}_i$ can also be written as

\begin{equation} \label{eqn:tetrad_scalars}
  \var \phi_i = \var F_{ab} \begin{cases}
    l^a m^b & i = 0 \\
    \frac{1}{2} (l^a n^b + \bar{m}^a m^b) & i = 1 \\
    \bar{m}^a n^b & i = 2
  \end{cases}, \qquad \var \bar{\chi}_i = \var F_{ab} \begin{cases}
    l^a \bar{m}^b & i = 0 \\
    \frac{1}{2} (l^a n^b + m^a \bar{m}^b) & i = 1 \\
    m^a n^b & i = 2
  \end{cases}.
\end{equation}
In Newman-Penrose notation, equation~\eqref{eqn:maxwell_spinor} becomes

\begin{subequations}
  \begin{align}
    (D - 2\rho) \var \phi_1 - (\bar{\delta} + \pi - 2\alpha) \var \phi_0 &= 2\pi \var J_l, \\
    (\delta - 2\tau) \var \phi_1 - (\npDelta + \mu - 2\gamma) \var \phi_0 &= 2\pi \var J_m, \\
    (D - \rho + 2\epsilon) \var \phi_2 - (\bar{\delta} + 2\pi) \var \phi_1 &= 2\pi \var J_{\bar{m}}, \\
    (\delta - \tau + 2\beta) \var \phi_2 - (\npDelta + 2\mu) \var \phi_1 &= 2\pi \var J_n,
  \end{align}
\end{subequations}
where, for any vector $v^a$, we define $v_l \equiv v^a l_a$, $v_n \equiv v^a n_a$, etc.

\subsection{Teukolsky formalism} \label{sec:teukolsky}

Next, we turn to the Teukolsky formalism.
The key to the Teukolsky formalism is that, in Kerr, there exists a \emph{Killing spinor} $\zeta_{AB}$, which is symmetric and satisfies the Killing spinor equation~\cite{penrose1988spinors}

\begin{equation}
  \nabla^{A'}{}_{(A} \zeta_{BC)} = 0.
\end{equation}
Consider a principal spin basis $(o, \iota)$, which is a spin basis whose associated null tetrad has the property (in Kerr) that the only non-zero component of the Weyl tensor $C_{abcd}$ is given by

\begin{equation}
  \Psi_2 \equiv C_{abcd} l^a m^b \bar{m}^c n^d.
\end{equation}
In this spin basis, there is also a scalar $\zeta$ such that

\begin{equation} \label{eqn:zeta_def}
  \zeta_{AB} = \zeta o_{(A} \iota_{B)},
\end{equation}
and $\zeta \sqrt[3]{\Psi_2}$ is constant (see~\cite{1970CMaPh..18..265W} for a proof, and more details).
Note that, given $\zeta_{AB}$, $\zeta$ can be determined up to a sign by

\begin{equation} \label{eqn:zeta_scalar}
  \zeta^2 = -2 \zeta_{AB} \zeta^{AB},
\end{equation}
but $\zeta_{AB}$ is only defined up to an overall constant.
The Killing spinor defines the Killing-Yano tensor $f_{ab}$ via~\cite{1970CMaPh..18..265W}

\begin{equation}
  f_{AA'BB'} = i\epsilon_{A'B'} \zeta_{AB} - i\epsilon_{AB} \bar{\zeta}_{A'B'},
\end{equation}
which can be used to define the Killing tensor $K_{ab}$ via

\begin{equation}
  K_{ab} = f_{ac} f^c{}_b.
\end{equation}

In terms of the scalar $\zeta$ and the principal spin basis $(o, \iota)$, we can define the \emph{master variables} $\pb{s} \Omega$ by~\cite{1973ApJ...185..635T}

\begin{equation} \label{eqn:decoupled_vars}
  \pb{s} \Omega = \begin{cases}
    \zeta^2 \var \phi_2 & s = -1 \\
    \var \phi_0 & s = 1
  \end{cases}.
\end{equation}
These master variables can also be written in terms of an operator acting on the vector potential: for $|s| = 1$,

\begin{equation} \label{eqn:defM}
  \pb{s} M_a \var A^a \equiv \pb{s} \Omega,
\end{equation}
where

\begin{equation} \label{eqn:M}
  \pb{s} M^a = \begin{cases}
    (D - \epsilon + \bar{\epsilon} - \bar{\rho}) m^a - (\delta + \bar{\pi} - \beta - \bar{\alpha}) l^a & s = 1 \\
    \zeta^2 [(\bar{\delta} + \alpha + \bar{\beta} - \bar{\tau}) n^a - (\mathbbold{\Delta} + \bar{\mu} + \gamma - \bar{\gamma}) \bar{m}^a] & s = -1
  \end{cases}.
\end{equation}

The master variables satisfy the \emph{Teukolsky equation}:

\begin{equation} \label{eqn:teukolsky}
  \pb{s} \Box \pb{s} \Omega = 8\pi \pb{s} \boldsymbol{\tau} \cdot \pb{|s|} \boldsymbol{T}.
\end{equation}
The operator on the left-hand side, $\pb{s} \Box$, is a second-order differential operator (the Teukolsky operator) given by, for $s \geq 0$~\cite{1973ApJ...185..635T}

\begin{subequations}
  \begin{align} \label{eqn:teukolsky_operator}
    &\phantom{_-} \begin{aligned}
      \pb{s} \Box = 2 \{&[D - (2s - 1) \epsilon + \bar{\epsilon} - 2s\rho - \bar{\rho}] (\npDelta - 2s\gamma + \mu) - [\npdelta - \bar{\alpha} - (2s - 1) \beta - 2s\tau + \bar{\pi}] (\bar{\npdelta} - 2s\alpha + \pi) \\
      &- 2 (2s - 1) (s - 1) \Psi_2\},
    \end{aligned} \\
    &\begin{aligned}
      \pb{-s} \Box = 2 \{&[\npDelta + (2s - 1) \gamma - \bar{\gamma} + \bar{\mu}] [D + 2s\epsilon + (2s - 1) \rho] - [\bar{\npdelta} + (2s - 1) \alpha + \bar{\beta} - \bar{\tau}] [\npdelta + 2s\beta + (2s - 1) \tau] \\
      &- 2 (2s - 1) (s - 1) \Psi_2\}.
    \end{aligned}
  \end{align}
\end{subequations}
On the right-hand side of the Teukolsky equation~\eqref{eqn:teukolsky} is the source term.
For $|s| = 1$, we have that $\pb{1} T_a = \var J_a$ and, reading off from equations (3.6) and (3.8) of~\cite{1973ApJ...185..635T}, we have\footnote{Note that these expressions for $\pb{s} \tau^a$ are not unique, since $\pb{s} \tau^a$ acts on the space of divergenceless vector fields.
  That is, the Teukolsky equation~\eqref{eqn:teukolsky} is preserved under the transformation $\pb{s} \tau^a \to \pb{s} \tau^a + \lambda \nabla^a$.}

\begin{equation} \label{eqn:tau}
  \pb{s} \tau^a = \frac{1}{2} \begin{cases}
    (\delta - \beta - \bar{\alpha} - 2\tau + \bar{\pi}) l^a - (D - \epsilon + \bar{\epsilon} - 2\rho - \bar{\rho}) m^a & s = 1 \\
    \zeta^2 [(\mathbbold{\Delta} + \gamma - \bar{\gamma} + 2\mu + \bar{\mu}) \bar{m}^a - (\bar{\delta} + \alpha + \bar{\beta} + 2\pi - \bar{\tau}) n^a] & s = -1
  \end{cases}.
\end{equation}
One can show that the procedure in~\cite{1973ApJ...185..635T} that was used to derive the Teukolsky equation~\eqref{eqn:teukolsky} is equivalent to the operator equation~\cite{PhysRevLett.41.203}

\begin{equation} \label{eqn:decoupling}
  \pb{s} \Box \pb{s} \boldsymbol{M} = \pb{s} \boldsymbol{\tau} \cdot \pb{|s|} \boldsymbol{\mathcal{E}}.
\end{equation}

In addition to decoupling the equations of motion for $\phi_0$ and $\phi_2$ from those for $\phi_1$, the Teukolsky formalism has the advantage that it results in separation of variables in a particular coordinate system and choice of null tetrad~\cite{1973ApJ...185..635T}.
The coordinate system is Boyer-Lindquist coordinates $\{t, r, \theta, \phi\}$, where the metric satisfies

\begin{equation}
  \ud s^2 = \ud t^2 - \Sigma \left(\frac{\ud r^2}{\Delta} + \ud \theta^2\right) - (r^2 + a^2) \sin^2 \theta \ud \phi^2 - \frac{2Mr}{\Sigma} \left(a \sin^2 \theta \ud \phi - \ud t\right)^2,
\end{equation}
where $\Delta = r^2 - 2Mr + a^2$ and $\Sigma = r^2 + a^2 \cos^2 \theta = |\zeta|^2$, where we have normalized $\zeta_{AB}$ such that

\begin{equation} \label{eqn:zeta_bl}
  \zeta = r - ia\cos \theta.
\end{equation}
The null tetrad associated with the principal spin basis in which the Teukolsky equation separates is the Kinnersley tetrad, defined by~\cite{1973ApJ...185..635T}

\begin{gather}
  \vec{l} = \frac{r^2 + a^2}{\Delta} \partial_t + \partial_r + \frac{a}{\Delta} \partial_\phi, \quad \vec{n} = \frac{1}{2\Sigma} \left[(r^2 + a^2) \partial_t - \Delta \partial_r + a \partial_\phi\right], \\
  \vec{m} = \frac{1}{\sqrt{2} \bar{\zeta}} \left(ia \sin \theta \partial_t + \partial_\theta + \frac{i}{\sin \theta} \partial_\phi\right).
\end{gather}
This spin basis has non-zero spin coefficients

\begin{gather}
  \rho = -\frac{1}{\zeta}, \quad \mu = -\frac{\Delta}{2\Sigma \zeta}, \quad \gamma = \mu + \frac{r - M}{2\Sigma}, \\
  \beta = \frac{\cot \theta}{2\sqrt{2} \bar{\zeta}}, \quad \pi = \alpha + \bar{\beta} = \frac{ia}{\sqrt{2} \zeta^2} \sin \theta, \quad \tau = -\frac{ia}{\sqrt{2} \Sigma} \sin \theta.
\end{gather}
We define the following operators~\cite{chandrasekhar1983mathematical, 1974ApJ...193..443T}, for integral parameters $n$ and $s$:

\begin{equation}
  \mathscr{D}_n = \partial_r + \frac{r^2 + a^2}{\Delta} \partial_t + \frac{a}{\Delta} \partial_\phi + 2n \frac{r - M}{\Delta}, \quad \mathscr{L}_s = \partial_\theta - i\left(a \sin \theta \partial_t + \frac{1}{\sin \theta} \partial_\phi\right) + s \cot \theta.
\end{equation}
We also define the operators $\mathscr{D}_n^+$ and $\mathscr{L}_n^+$, where the ``+'' operation is defined by taking $\partial_t \to -\partial_t$ and $\partial_\phi \to -\partial_\phi$, and so it follows that $\mathscr{L}_s^+ = \overline{\mathscr{L}_s}$.\footnote{As with $\delta$, we take $\overline{\mathscr{L}_s} f$ to mean $\overline{\mathscr{L}_s \bar{f}}$, not $\overline{\mathscr{L}_s f}$.}
These operators act on Fourier modes $f_{m\omega} e^{i(m\phi - \omega t)}$ to yield the operators $\mathscr{D}_{nm\omega}$ and $\mathscr{L}_{sm\omega}$:

\begin{equation}
  \mathscr{D}_n \left[f_{m\omega} e^{i(m\phi - \omega t)}\right] = e^{i(m\phi - \omega t)} \mathscr{D}_{nm\omega} f_{m\omega}, \quad \mathscr{L}_s \left[f_{m\omega} e^{i(m\phi - \omega t)}\right] = e^{i(m\phi - \omega t)} \mathscr{L}_{sm\omega} f_{m\omega},
\end{equation}
where

\begin{equation} \label{eqn:mode_DL}
  \mathscr{D}_{nm\omega} \equiv \partial_r + \frac{iK_{m\omega}}{\Delta} + 2n \frac{r - M}{\Delta}, \quad \mathscr{L}_{sm\omega} \equiv \partial_\theta + Q_{m\omega} + s \cot \theta,
\end{equation}
and

\begin{equation}
  K_{m\omega} \equiv am - \omega (r^2 + a^2), \quad Q_{m\omega} \equiv m \csc \theta - a \omega \sin \theta.
\end{equation}
Note that we follow the sign convention for $K_{m\omega}$ of Chandrasekhar~\cite{chandrasekhar1983mathematical}.

We now consider the Teukolsky equation~\eqref{eqn:teukolsky} in Boyer-Lindquist coordinates.
Define the following Fourier modes $\pb{s} \widetilde{\Omega}_{m\omega} (r, \theta)$ for $\pb{s} \Omega$ by

\begin{equation}
  \pb{s} \Omega(t, r, \theta, \phi) = \int_{-\infty}^\infty \ud \omega \sum_{m = 0}^\infty e^{i(m\phi - \omega t)} \pb{s} \widetilde{\Omega}_{m\omega} (r, \theta).
\end{equation}
In terms of these modes and the operators in equation~\eqref{eqn:mode_DL}, the Teukolsky equation~\eqref{eqn:teukolsky} (without sources) becomes~\cite{1973ApJ...185..635T}

\begin{equation}
  \left[\Delta \mathscr{D}_{(1 - s) \pm m \pm \omega} \mathscr{D}_{0 \mp m \mp \omega} + \mathscr{L}_{(1 - s) \mp m \mp \omega} \mathscr{L}_{s \pm m \pm \omega} \pm 2 (2s - 1) i\omega (r - ia \cos \theta)\right] \Delta^{(s \pm s)/2} \pb{\pm s} \widetilde{\Omega}_{m\omega} (r, \theta) = 0
\end{equation}
(for $s \geq 0$).
This equation separates in $r$ and $\theta$, and so one can write the following expansion:

\begin{equation} \label{eqn:decoupled_modes}
  \pb{s} \Omega(t, r, \theta, \phi) = \int_{-\infty}^\infty \ud \omega \sum_{l = |s|}^\infty \sum_{|m| \leq l} e^{i(m\phi - \omega t)} \pb{s} \Theta_{lm\omega} (\theta) \pb{s} \widehat{\Omega}_{lm\omega} (r),
\end{equation}
and upon substituting this expansion into the Teukolsky equation~\eqref{eqn:teukolsky} (again without sources) one finds that

\begin{subequations}
  \begin{align}
    \left[\mathscr{L}_{(1 - s) \mp m \mp \omega} \mathscr{L}_{s \pm m \pm \omega} \pm 2 (2s - 1) \omega a \cos \theta + \pb{\pm s} \lambda_{lm\omega}\right] \pb{\pm s} \Theta_{lm\omega} (\theta) &= 0, \label{eqn:angular_teukolsky} \\
    \left[\Delta \mathscr{D}_{(1 - s) \pm m \pm \omega} \mathscr{D}_{0 \mp m \mp \omega} \pm 2 (2s - 1) i\omega r - \pb{\pm s} \lambda_{lm\omega}\right] \Delta^{(s \pm s)/2} \pb{\pm s} \widehat{\Omega}_{lm\omega} (r) &= 0 \label{eqn:radial_teukolsky}
  \end{align}
\end{subequations}
(for $s \geq 0$).
The separation constant $\pb{\pm s} \lambda_{lm\omega}$ reduces to $l(l + 1) - s(s - 1)$ in the Schwarzschild limit, and so $l(l + 1)$ for the electromagnetic case~\cite{1973ApJ...185..635T,chandrasekhar1983mathematical}.

The functions $\pb{s} \Theta_{lm\omega} (\theta)$ are solutions to an eigenvalue problem for a separation constant $\pb{s} \lambda_{lm\omega}$ that are regular on $[0, \pi]$.
We fix their definition by noting that the symmetries of equation~\eqref{eqn:angular_teukolsky} imply that we can choose these functions to be real, and also to satisfy

\begin{equation} \label{eqn:phase_conventions}
  \pb{s} \Theta_{lm\omega} (\theta) = (-1)^{m + s} \pb{-s} \Theta_{l(-m)(-\omega)} (\theta), \qquad \pb{s} \Theta_{lm\omega} (\pi - \theta) = (-1)^{l + m} \pb{-s} \Theta_{lm\omega} (\theta),
\end{equation}
with the separation constant also being real and satisfying

\begin{equation}
  \pb{s} \lambda_{lm\omega} = \pb{-s} \lambda_{lm\omega} = \pb{s} \lambda_{l(-m)(-\omega)}.
\end{equation}
For different $l$, the functions $\pb{s} \Theta_{lm\omega} (\theta)$ are orthogonal, and we normalize them by requiring that

\begin{equation} \label{eqn:angular_normalization}
  \int_0^\pi \ud \theta \pb{s} \Theta_{lm\omega} (\theta) \pb{s} \Theta_{l'm\omega} (\theta) \sin \theta \ud \theta = \delta_{ll'}.
\end{equation}
The functions $e^{i(m\phi - \omega t)} \pb{s} \Theta_{lm\omega} (\theta)$ are known as the \emph{spin-weighted spheroidal harmonics}, and are orthogonal for different $l$, $m$, and $\omega$.

The functions $\pb{s} \widehat{\Omega}_{lm\omega} (r)$ occur as coefficients of the expansion of $\pb{s} \Omega$ in the spin-weighted spheroidal harmonics.
Since they obey equation~\eqref{eqn:radial_teukolsky}, which is invariant under the operation of complex conjugation followed by $(m, \omega) \to (-m, -\omega)$, we can define two linearly independent solutions, labelled by their eigenvalue $p = \pm 1$ under this operation [multiplied by a conventional factor of $(-1)^{m + s}$]:

\begin{equation} \label{eqn:parity}
  \pb{s} \widehat{\Omega}_{lm\omega p} (r) = \frac{1}{2} \bigg[\pb{s} \widehat{\Omega}_{lm\omega} (r) + p(-1)^{m + s} \overline{\pb{s} \widehat{\Omega}_{l(-m)(-\omega)} (r)}\bigg].
\end{equation}
Depending on the situation, it may be convenient to expand the master variables either as in equation~\eqref{eqn:decoupled_modes}, or as

\begin{equation} \label{eqn:decoupled_modes_parity}
  \pb{s} \Omega(t, r, \theta, \phi) = \int_{-\infty}^\infty \ud \omega \sum_{l = |s|}^\infty \sum_{|m| \leq l} \sum_{p = \pm 1} e^{i(m\phi - \omega t)} \pb{s} \Theta_{lm\omega} (\theta) \pb{s} \widehat{\Omega}_{lm\omega p} (r).
\end{equation}
In particular, the complex conjugate of the master variables has a natural expansion in this form:

\begin{equation} \label{eqn:conjugate_modes_parity}
  \overline{\pb{s} \Omega(t, r, \theta, \phi)} = \int_{-\infty}^\infty \ud \omega \sum_{l = |s|}^\infty \sum_{|m| \leq l} \sum_{p = \pm 1} p e^{i(m\phi - \omega t)} \pb{-s} \Theta_{lm\omega} (\theta) \pb{s} \widehat{\Omega}_{lm\omega p} (r),
\end{equation}
from equations~\eqref{eqn:phase_conventions} and~\eqref{eqn:parity}.

\section{Symmetry operators} \label{sec:symmetry_operators}

A \emph{symmetry operator}, as defined by Kalnins, McLenaghan, and Williams~\cite{1992RSPSA.439..103K}, is an $\mathbb{R}$-linear operator that maps the space of solutions to the equations of motion into itself.
For complexified solutions to real equations of motion, for example, there exists a trivial symmetry operator defined by complex conjugation.

In~\cite{PhysRevD.16.3395}, Carter found a symmetry operator $\mathcal{D}$ that mapped the space of solutions to the scalar field equation to itself, given by

\begin{equation}
  \mathcal{D} \equiv \nabla_a K^{ab} \nabla_b.
\end{equation}
This symmetry operator explicitly depends on the Killing tensor in the Kerr spacetime.
The symmetry operators which we discuss in this paper are constructed, instead, from the Killing spinor $\zeta_{AB}$ that can be used to define the Killing tensor, as we did in section~\ref{sec:teukolsky}.

In the remainder of this section, we review three classes of symmetry operators, and review one particular consequence of symmetry operators in Kerr, the \emph{Teukolsky-Starobinsky identities}.

\subsection{Killing vectors} \label{sec:killing}

An example of a symmetry operator is given by the action of the Lie derivative by a Killing vector, since this operator commutes with all derivative operators in the equations of motion.
This operator is \emph{first-order}, in the sense that it contains only one derivative acting on the solution.
In the case of spin 1, there are, in fact, two symmetry operators defined by a Killing vector $\xi^a$: the first is just the normal Lie derivative $\lie_{\xi}$, and the second (defined by Andersson, B\"ackdahl, and Blue~\cite{Andersson:2015xla}) is the mapping

\begin{equation} \label{eqn:def_xiL}
  \pb{\xi} L_a{}^b \var A_b \equiv \xi_{A'}{}^B \pb{1} \mathcal{F}_{AB}{}^c \var A_c,
\end{equation}
where $\pb{1} \mathcal{F}_{AB}{}^c$ is defined in equation~\eqref{eqn:faraday_spinor}.
The latter is the first example of a \emph{gauge-invariant} symmetry operator, in the sense that

\begin{equation}
  \pb{\xi} \boldsymbol{L} \cdot \ud \lambda = 0,
\end{equation}
while the former is a \emph{gauge-covariant} symmetry operator, since

\begin{equation}
  \lie_{\xi} \ud \lambda = \ud \lie_{\xi} \lambda,
\end{equation}
which is still exact.
A gauge-invariant symmetry operator has fixed the gauge, whereas a gauge-covariant symmetry operator still retains gauge freedom through the original vector potential.

Consider a symmetry operator which is constructed from $\pb{1} \boldsymbol{\mathcal{F}}$, like $\pb{\xi} \boldsymbol{L}$ in equation~\eqref{eqn:def_xiL}.
Such a symmetry operator may map $\var \boldsymbol{A}$ to a vector potential which has an entirely anti-self-dual or self-dual exterior derivative.
Such symmetry operators are then considered to be of the \emph{first} or \emph{second kind}, respectively.
For example, $\pb{\xi} \boldsymbol{L}$ is of the first kind.
This can be seen as follows:

\begin{equation}
  \overline{\pb{1} \mathcal{F}_{A'B'}{}^c} \pb{\xi} L_c{}^d \var A_d = \nabla^A{}_{(A'} \left(\xi_{B')}{}^B \pb{1} \mathcal{F}_{AB}{}^c \var A_c\right) = \left(\nabla^{(A}{}_{(A'} \xi_{B')}{}^{B)}\right) \pb{1} \mathcal{F}_{AB}{}^c \var A_c,
\end{equation}
which vanishes since $\nabla_{(a} \xi_{b)} = 0$.
Note that, in spin 1 (with no sources), any mapping to a vector field with a(n) (anti-)self-dual exterior derivative is a symmetry operator: if $\dual \ud \var \boldsymbol{A} = \pm i \ud \var \boldsymbol{A}$, then

\begin{equation}
  \ud \dual \ud \var \boldsymbol{A} = \pm i \ud^2 \var \boldsymbol{A} = 0.
\end{equation}
and so Maxwell's equations are automatically satisfied.
This completes the proof that $\pb{\xi} \boldsymbol{L}$ is a symmetry operator.

Higher-order symmetry operators can be created by composing these first-order symmetry operators, but they are reducible, in the same sense that Killing tensors that are products of Killing vectors are reducible.
In Kerr, there are many higher-order symmetry operators that are irreducible, which we will discuss in Sections~\ref{sec:separation} and~\ref{sec:tsi}.

\subsection{First kind} \label{sec:separation}

Another class of symmetry operators is given by separation of variables.
The separability of a differential equation in two variables ($r$ and $\theta$ in the case of the Teukolsky equation in Kerr) implies that there are two operators whose eigenfunctions span the space of solutions.
As an explicit example, consider the case of a separable differential equation in two variables $x_1$ and $x_2$:

\begin{equation} \label{eqn:sep_example}
  \left[\mathcal{L}_1 (x_1, \partial_{x_1}, \ldots) + \mathcal{L}_2 (x_2, \partial_{x_2}, \ldots)\right] P(x_1, x_2) = 0.
\end{equation}
By the separability of this equation, one can write a general solution as

\begin{equation}
  P(x_1, x_2) = \sum_\lambda P_1^\lambda (x_1) P_2^\lambda (x_2),
\end{equation}
where

\begin{equation} \label{eqn:sep_constants}
  \mathcal{L}_1 P_1^\lambda (x_1) = \lambda P_1^\lambda (x_1), \qquad \mathcal{L}_2 P_2^\lambda (x_2) = -\lambda P_2^\lambda (x_2).
\end{equation}
The operators $\mathcal{L}_1$ and $\mathcal{L}_2$ act as symmetry operators, since

\begin{equation}
  \mathcal{L}_1 P(x_1, x_2) = \sum_\lambda \lambda P_1^\lambda (x_1) P_2^\lambda (x_2), \qquad \mathcal{L}_2 P(x_1, x_2) = -\sum_\lambda \lambda P_1^\lambda (x_1) P_2^\lambda (x_2),
\end{equation}
which are both separately solutions to~\eqref{eqn:sep_example}, using equations~\eqref{eqn:sep_constants}.

Since the Teukolsky equation~\eqref{eqn:teukolsky} separates in $r$ and $\theta$, there is a symmetry operator on the space of master variables associated with this separation.
As shown in~\cite{Kalnins:jmp/30/10/10.1063/1.528565}, there exists a symmetry operator of the first kind $\Lambda_{AB}{}^{CD}$ that maps the space of spinor fields $\var \phi_{AB}$ satisfying equation~\eqref{eqn:maxwell_spinor} with $J^{AA'} = 0$ to itself, which corresponds to the symmetry operator on the space of master variables derived from separation of variables in $r$ and $\theta$.
However, this is a symmetry operator for the spinor field; in this paper, we are concerned with symmetry operators for the space of vector potentials, as those are the ones that we will use with the symplectic product in Section~\ref{sec:symplectic} below.
A second-order symmetry operator of the first kind, for the vector potential, was defined by~\cite{Andersson:2014lca} as

\begin{equation} \label{eqn:def_A}
  \mathcal{A}_{AA'}{}^b \var A_b = -\tfrac{1}{3} \left(\nabla_{BB'} \bar{\zeta}_{A'}{}^{B'}\right) \pb{2} \mathcal{N}_A{}^{BCD} \pb{1} \mathcal{F}_{CD}{}^e \var A_e + \bar{\zeta}_{A'}{}^{B'} \nabla_{BB'} \left(\pb{2} \mathcal{N}_A{}^{BCD} \pb{1} \mathcal{F}_{CD}{}^e \var A_e\right),
\end{equation}
where we have defined, for any symmetric spinor field $S_{AB}$,

\begin{equation} \label{eqn:def_N}
  \pb{2} \mathcal{N}_{AB}{}^{CD} S_{CD} \equiv \zeta_{(A}{}^C S_{B)C}.
\end{equation}
Since $o^A \iota^B \pb{2} \mathcal{N}_{AB}{}^{CD} S_{CD} = 0$ by equation~\eqref{eqn:zeta_def}, $\pb{2} \boldsymbol{\mathcal{N}}$ sets to zero the middle scalar $S_1$ of $S_{AB}$ [from equation~\eqref{eqn:scalar_def}].

\subsection{Second kind} \label{sec:tsi}

Next, we move on to the symmetry operators which correspond to the Teukolsky-Starobinsky identities~\cite{PhysRevD.10.1070, PhysRevLett.41.203, 1985RSPSA.400..119C}.
These are the symmetry operators which most clearly generalize to the case of linearized gravity.
There are two approaches to deriving these symmetry operators: either one starts by writing down the most general symmetry operator, as was done in~\cite{Andersson:2014lca} to derive the results in Section~\ref{sec:separation}, or one uses the properties of the operator Teukolsky formalism, as presented in Section~\ref{sec:teukolsky}.

Using the approach of deriving the most generic operator, which works for spin 1/2~\cite{PhysRevD.19.1093}, a symmetry operator of the second kind, mapping between spinor fields, was found by Kalnins, Miller, and Williams~\cite{Kalnins:jmp/30/10/10.1063/1.528565} (and later corrected in~\cite{1992RSPSA.439..103K}).
Similarly, Andersson, B\"ackdahl, and Blue wrote down the most generic second-order symmetry operator of the second kind for vector potentials in Kerr~\cite{Andersson:2014lca}:

\begin{equation} \label{eqn:B}
  \mathcal{B}_{AA'}{}^b \var A_b = \tfrac{1}{3} \left(\nabla_{CA'} \zeta_B{}^C\right) \pb{2} \mathcal{N}_A{}^{BCD} \pb{1} \mathcal{F}_{CD}{}^e \var A_e + \zeta_A{}^B \nabla_{CA'} \left(\pb{2} \mathcal{N}_B{}^{CDE} \pb{1} \mathcal{F}_{DE}{}^f \var A_f\right).
\end{equation}
Torres del Castillo~\cite{1985RSPSA.400..119C}, similarly, found a mapping between vector potentials of the form

\begin{equation} \label{eqn:T}
  \pb{1} \mathcal{T}_{AA'}{}^b \var A_b = \zeta^{-2} \nabla^B{}_{A'} \left(\zeta^2 \zeta_{AC} \zeta_{BD} \pb{1} \mathcal{F}^{CD}{}_e \var A^e\right),
\end{equation}
which is a symmetry operator in algebraically general spacetimes as well.
However, these symmetry operators do not obviously generalize to spin 2, since the proofs of their validity depend crucially on using equation~\eqref{eqn:maxwell_spinor}, which does not have a clean (and gauge-invariant) generalization to spin 2.

An approach that does have a clear generalization is due to Wald, which relies on the operator Teukolsky formalism~\cite{PhysRevLett.41.203}.
We define the following notion of the adjoint: for tensor fields $\boldsymbol{\phi}$ and $\boldsymbol{\psi}$ of rank $p$ and $q$, respectively, and any linear differential operator $\boldsymbol{\mathcal{L}}$ that takes tensor fields of rank $p$ to $q$, then we say that $\boldsymbol{\mathcal{L}}^\dagger$ is the adjoint of $\boldsymbol{\mathcal{L}}$ if there exists a vector field $j^a [\boldsymbol{\phi}, \boldsymbol{\psi}]$ such that\footnote{If there is no restriction on the domain of $\boldsymbol{\mathcal{L}}$, the operator $\boldsymbol{\mathcal{L}}$ which satisfies equation~\eqref{eqn:def_adjoint} is unique.}

\begin{equation} \label{eqn:def_adjoint}
  \boldsymbol{\psi} \cdot (\mathcal{L} \cdot \boldsymbol{\phi}) - \boldsymbol{\phi} \cdot (\mathcal{L}^\dagger \cdot \boldsymbol{\psi}) \equiv \nabla_a j^a [\boldsymbol{\phi}, \boldsymbol{\psi}].
\end{equation}
With this definition, $\pb{s} \boldsymbol{\mathcal{E}}$ is self-adjoint, and so we can take the adjoint of equation~\eqref{eqn:decoupling} to obtain

\begin{equation} \label{eqn:adjoint_decoupling}
  \pb{|s|} \boldsymbol{\mathcal{E}} \cdot \pb{s} \boldsymbol{\tau}^\dagger = \pb{s} \boldsymbol{M}^\dagger \pb{s} \Box^\dagger.
\end{equation}
Applying this new operator equation to a solution to the adjoint equation $\pb{s} \Box^\dagger \Phi = 0$, we obtain

\begin{equation}
  \pb{|s|} \boldsymbol{\mathcal{E}} \cdot \pb{s} \boldsymbol{\tau}^\dagger \Phi = 0.
\end{equation}
Thus, for example, $\pb{\pm 1} \tau_a^\dagger \Phi$ is a vector potential that solves Maxwell's equations.
Using~\eqref{eqn:npadjoint} and~\eqref{eqn:tau}, we have that

\begin{equation} \label{eqn:tau_dagger}
  \pb{s} \tau_a^\dagger = \frac{1}{2} \begin{cases}
    -l_a (\delta + 2\beta + \tau) + m_a (D + 2\epsilon + \rho) & s = 1 \\
    [-\bar{m}_a (\mathbbold{\Delta} - \mu - 2\gamma) + n_a (\bar{\delta} - 2\alpha - \pi)] \zeta^2 & s = -1
  \end{cases}.
\end{equation}

Such a solution $\Phi$ that generates a vector potential is referred to as a \emph{Debye potential}.
As was first noted by Cohen and Kegeles~\cite{PhysRevD.10.1070}, $\pb{s} \Box^\dagger = \pb{-s} \Box$, which can be derived from equation~\eqref{eqn:def_adjoint} and

\begin{subequations}
  \begin{align} \label{eqn:npadjoint}
    D^\dagger &= -D - (\epsilon + \bar{\epsilon}) + \rho + \bar{\rho}, \\
    \mathbbold{\Delta}^\dagger &= -\mathbbold{\Delta} + \gamma + \bar{\gamma} - (\mu + \bar{\mu}), \\
    \delta^\dagger &= -\delta + \bar{\alpha} - \beta + \tau - \bar{\pi}.
  \end{align}
\end{subequations}
Therefore, one can use solutions to the vacuum Teukolsky equation~\eqref{eqn:teukolsky} $\pb{-s} \Box \pb{-s} \Phi = 0$ to get new solutions, since from equations~\eqref{eqn:adjoint_decoupling} and~\eqref{eqn:decoupling},

\begin{equation} \label{eqn:tsi}
  \pb{\pm s} \Box \pb{\pm s} \boldsymbol{M} \cdot \pb{s} \boldsymbol{\tau}^\dagger \pb{-s} \Phi = 0, \qquad \pb{\pm s} \Box \pb{\pm s} \boldsymbol{M} \cdot \overline{\pb{s} \boldsymbol{\tau}^\dagger \pb{-s} \Phi} = 0.
\end{equation}
These are sometimes called the \emph{Teukolsky-Starobinsky identities}, and we discuss them further in Section~\ref{sec:action}.
Thus, there exist two operators $\pb{\pm S} \boldsymbol{M} \cdot \pb{-S} \boldsymbol{\tau}^\dagger$ which map the space of master variables with $s = -S$ to the space of master variables with $s = \pm S$, and two operators $\pb{\pm S} \boldsymbol{M} \cdot \overline{\pb{-S} \boldsymbol{\tau}^\dagger}$ which map the space of complex conjugated master variables with $s = S$ to the space of master variables $S = \pm s$.

We can also form the following operators that act on the vector potential (where $|s| = 1$):

\begin{equation} \label{eqn:C_def}
  \pb{s} \mathcal{C}_a{}^b \equiv \pb{s} \tau_a^\dagger \pb{-s} M^b.
\end{equation}
If $\var A_a$ is a source-free solution to Maxwell's equations, then from equations~\eqref{eqn:C_def},~\eqref{eqn:adjoint_decoupling}, and~\eqref{eqn:decoupling}, we have

\begin{equation}
  \pb{|s|} \boldsymbol{\mathcal E} \cdot \pb{s} \boldsymbol{\mathcal C} \cdot \var \boldsymbol{A} = 0.
\end{equation}
These operators therefore act as symmetry operators for the vector potential in Kerr.

The operators $\pb{\pm 1} \boldsymbol{\mathcal{C}}$ are symmetry operators of the second kind, since for a source-free solution $\var A_a$ of Maxwell's equations, the only non-zero combinations of $\pb{1} \boldsymbol{\mathcal{F}}$ and $\overline{\pb{1} \boldsymbol{\mathcal{F}}}$ acting on $\pb{\pm 1} \boldsymbol{\mathcal{C}} \cdot \var \boldsymbol{A}$ are

\begin{subequations} \label{eqn:2nd_kind}
  \begin{align}
    \left(\overline{\pb{1} \boldsymbol{\mathcal{F}}} \cdot \pb{s} \boldsymbol{\mathcal{C}} \cdot \var \boldsymbol{A}\right)_0 &= -\frac{1}{2} \begin{cases}
      (D + \epsilon - \bar{\epsilon} - \rho) (D + 2\epsilon + \rho) \zeta^2 \var \phi_2 & s = 1 \\
      (\bar{\delta} - \bar{\beta} - \alpha + \bar{\pi})(\bar{\delta} - 2\alpha - \pi) \zeta^2 \var \phi_0 & s = -1
    \end{cases}, \label{eqn:2nd_kind_0}\\
    \left(\overline{\pb{1} \boldsymbol{\mathcal{F}}} \cdot \pb{s} \boldsymbol{\mathcal{C}} \cdot \var \boldsymbol{A}\right)_1 &= -\frac{1}{4} \begin{cases}
      \begin{aligned}
        [(D + \epsilon + \bar{\epsilon} - \rho + \bar{\rho}) &(\delta + 2\beta + \tau) \\
        &+ (\delta + \beta - \bar{\alpha} - \bar{\pi} - \tau) (D + 2\epsilon + \rho)] \zeta^2 \var \phi_2
      \end{aligned} & s = 1 \\
      \begin{aligned}
        [(\mathbbold{\Delta} - \gamma - \bar{\gamma} + \mu - \bar{\mu}) &(\bar{\delta} - 2\alpha - \pi) \\
        &+ (\bar{\delta} - \alpha + \bar{\beta} + \bar{\tau} + \pi) (\mathbbold{\Delta} - 2\gamma - \mu)] \zeta^2 \var \phi_0
      \end{aligned} & s = -1
    \end{cases}, \label{eqn:2nd_kind_1} \\
    \left(\overline{\pb{1} \boldsymbol{\mathcal{F}}} \cdot \pb{s} \boldsymbol{\mathcal{C}} \cdot \var \boldsymbol{A}\right)_2 &= -\frac{1}{2} \begin{cases}
      (\delta + \bar{\alpha} + \beta - \tau) (\delta + 2\beta + \tau) \zeta^2 \var \phi_2 & s = 1 \\
      (\mathbbold{\Delta} - \gamma + \bar{\gamma} + \mu) (\mathbbold{\Delta} - 2\gamma - \mu) \zeta^2 \var \phi_0 & s = -1
    \end{cases}. \label{eqn:2nd_kind_2}
  \end{align}
\end{subequations}
Note that the $s = 1$ version of equation~\eqref{eqn:2nd_kind_0} is the same as the $s = -1$ version of equation~\eqref{eqn:2nd_kind_2}, up to a transformation of the form $l^a \leftrightarrow n^a$ and $m^a \leftrightarrow \bar{m}^a$.
This same transformation gives the $s = -1$ version of equation~\eqref{eqn:2nd_kind_0} from the $s = 1$ version of equation~\eqref{eqn:2nd_kind_2}, as well as the $s = -1$ version of equation~\eqref{eqn:2nd_kind_1} from the $s = 1$ version of equation~\eqref{eqn:2nd_kind_1}.

\subsection{Issues of gauge}

The argument given in Section~\ref{sec:tsi} implies that there exist two symmetry operators of the second kind, given by $\pb{\pm 1} \boldsymbol{\mathcal{C}}$.
Similarly, the existence of symmetry operators of the second kind result in the four Teukolsky-Starobinsky identities [the second equation in~\eqref{eqn:tsi}, which is four equations because of the two choices of $s$ and the two choices of $\pm$].
However, the four Teukolsky-Starobinsky identities are in fact related, which imply that $\pb{\pm 1} \boldsymbol{\mathcal{C}} \cdot \var \boldsymbol{A}$ are related by a gauge transformation.
To see this, note that Maxwell's equations in a vacuum can be written in the form

\begin{subequations}
  \begin{align}
    D \zeta^2 \var \phi_1 &= (\bar{\delta} - 2\alpha - \pi) \zeta^2 \var \phi_0, \\
    \delta \zeta^2 \var \phi_1 &= (\mathbbold{\Delta} - 2\gamma - \mu) \zeta^2 \var \phi_0, \\
    \bar{\delta} \zeta^2 \var \phi_1 &= (D + 2\epsilon + \rho) \zeta^2 \var \phi_2, \\
    \mathbbold{\Delta} \zeta^2 \var \phi_1 &= (\delta + 2\beta + \tau) \zeta^2 \var \phi_2,
  \end{align}
\end{subequations}
so that, using the Newman-Penrose commutators (see~\cite{penrose1987spinors}),

\begin{subequations} \label{eqn:tsr}
  \begin{align}
    \left(\overline{\pb{1} \boldsymbol{\mathcal{F}}} \cdot (\pb{+1} \boldsymbol{\mathcal{C}} - \pb{-1} \boldsymbol{\mathcal{C}}) \cdot \var \boldsymbol{A}\right)_0 &= -\frac{1}{2} [(D + \epsilon - \bar{\epsilon} - \rho) \bar{\delta} - (\bar{\delta} - \bar{\beta} - \alpha + \bar{\pi}) D] \zeta^2 \var \phi_1 = 0, \\
    \left(\overline{\pb{1} \boldsymbol{\mathcal{F}}} \cdot (\pb{+1} \boldsymbol{\mathcal{C}} - \pb{-1} \boldsymbol{\mathcal{C}}) \cdot \var \boldsymbol{A}\right)_1 &= -\frac{1}{4} [(D + \epsilon + \bar{\epsilon} - \rho - \bar{\rho}) \mathbbold{\Delta} + (\delta + \beta - \bar{\alpha} - \bar{\pi} - \tau) \bar{\delta} \nonumber \\
                                                                                                                                          &\qquad\quad- (\mathbbold{\Delta} - \gamma - \bar{\gamma} + \mu - \bar{\mu}) D - (\bar{\delta} - \alpha + \bar{\beta} + \bar{\tau} + \pi) \delta] \zeta^2 \var \phi_1 = 0, \\
    \left(\overline{\pb{1} \boldsymbol{\mathcal{F}}} \cdot (\pb{+1} \boldsymbol{\mathcal{C}} - \pb{-1} \boldsymbol{\mathcal{C}}) \cdot \var \boldsymbol{A}\right)_2 &= -\frac{1}{2} [(\delta + \bar{\alpha} + \beta - \tau) \mathbbold{\Delta} - (\mathbbold{\Delta} - \gamma + \bar{\gamma} + \mu) \delta] \zeta^2 \var \phi_1 = 0.
  \end{align}
\end{subequations}
Note that these equations hold both for real and complexified solutions to Maxwell's equations.
These equations mean that the two different operators $\pb{1} \boldsymbol{\mathcal{C}}$ and $\pb{-1} \boldsymbol{\mathcal{C}}$ are equivalent up to gauge.
Furthermore, since equations~\eqref{eqn:2nd_kind} agree with the results of Torres del Castillo (up to constant multiplicative factors)~\cite{1985RSPSA.400..119C}, our operators $\pb{\pm 1} \mathcal{C}$ are the same as $\mathcal{T}$ in equation~\eqref{eqn:T}, up to constant factors and gauge.

\subsection{Teukolsky-Starobinsky identities} \label{sec:action}

In this section, we show that the symmetry operators in Section~\ref{sec:tsi} are in a sense ``diagonal'', multiplying each term in the expansion~\eqref{eqn:conjugate_modes_parity} of the complex conjugated master variables by a constant.
Consider the symmetry operators $\pb{\pm 1} M^a \overline{\pb{1} \tau_a^\dagger}$ and $\pb{\pm 1} M^a \overline{\pb{-1} \tau_a^\dagger}$ that [from equation~\eqref{eqn:tsi}] map the space of complex conjugated master variables with $s = -1$ and $s = 1$, respectively, to the space of master variables with $s = \pm 1$.
One can show that these symmetry operators can be written in terms of purely radial and purely angular operators, the so called ``spin-inversion'' operators:

\begin{subequations}
  \begin{align}
    \pb{1} M^a \overline{\pb{1} \tau_a^\dagger} = -\frac{1}{2} \mathscr{D}_0^2, &\qquad \pb{-1} M^a \overline{\pb{-1} \tau_a^\dagger} = -\frac{1}{8} \Delta \left(\mathscr{D}_0^+\right)^2 \Delta, \label{eqn:radial_spin_inv} \\
    \pb{1} M^a \overline{\pb{-1} \tau_a^\dagger} = -\frac{1}{4} \mathscr{L}_0^+ \mathscr{L}_1^+, &\qquad \pb{-1} M^a \overline{\pb{1} \tau_a^\dagger} = -\frac{1}{4} \mathscr{L}_0 \mathscr{L}_1. \label{eqn:angular_spin_inv}
  \end{align}
\end{subequations}
The existence of these operators imply certain relationships between $\pb{\pm 1} \Theta_{lm\omega} (\theta)$, as well as between $\pb{\pm 1} \widehat{\Omega}_{lm\omega} (r)$, assuming that $\pb{\pm 1} \Omega$ are generated by the same vector potential via equations~\eqref{eqn:tetrad_scalars} and~\eqref{eqn:decoupled_vars}.

We start with the operators in equation~\eqref{eqn:angular_spin_inv}.
Since these operators are purely angular, equations~\eqref{eqn:conjugate_modes_parity} and~\eqref{eqn:tsi} imply that $\pb{1} M^a \overline{\pb{-1} \tau^\dagger}$ maps solutions to equation~\eqref{eqn:angular_teukolsky} for $s = -1$ to solutions for $s = 1$.
Similarly, $\pb{-1} M^a \overline{\pb{1} \tau^\dagger}$ maps solutions for $s = 1$ to solutions for $s = -1$.
Since the solutions $\pb{\pm s} \Theta_{lm\omega} (\theta)$, for a given $l$, $m$, $\omega$, and $s$, are fully determined (up to normalization) by the boundary conditions of regularity on $[0, \pi]$, we have that

\begin{equation} \label{eqn:angular_tsi}
  \mathscr{L}_{0(\pm m)(\pm \omega)} \mathscr{L}_{1(\pm m)(\pm \omega)} \pb{\pm 1} \Theta_{lm\omega} (\theta) \equiv \pb{\pm 1} B_{lm\omega} \pb{\mp 1} \Theta_{lm\omega} (\theta),
\end{equation}
for some constants $\pb{\pm 1} B_{lm\omega}$.
These constants are determined by the normalization of $\pb{\pm 1} \Theta_{lm\omega} (\theta)$ in equation~\eqref{eqn:angular_normalization}; one finds that~\cite{chandrasekhar1983mathematical}

\begin{equation}
  \pb{1} B_{lm\omega} = \pb{-1} B_{lm\omega} \equiv B_{lm\omega}, \qquad B_{lm\omega}^2 = \pb{1} \lambda_{lm\omega}^2 - 4\alpha_{m\omega}^2 \omega^2,
\end{equation}
where $\alpha_{m\omega}^2 = a^2 - am/\omega$.
The sign of $B_{lm\omega}$ is fixed by setting $a\omega = 0$, at which point equation~\eqref{eqn:angular_tsi} relates spin-weighted \emph{spherical} harmonics, and one finds that $B_{lm\omega} \geq 0$~\cite{1974ApJ...193..443T}.
Equation~\eqref{eqn:angular_tsi} is sometimes known as the \emph{angular} Teukolsky-Starobinsky identity, where $B_{lm\omega}$ is the \emph{Teukolsky-Starobinsky constant}.

Next, we consider the radial version of the Teukolsky-Starobinsky identity.
Here, we note that the operators in equation~\eqref{eqn:radial_spin_inv} are purely radial, and moreover (when acting on Fourier modes) invariant under complex conjugation composed with the transformation $(m, \omega) \to (-m, -\omega)$.
Therefore the operator $\pb{1} M^a \overline{\pb{1} \tau_a^\dagger}$ [by equations~\eqref{eqn:conjugate_modes_parity} and~\eqref{eqn:tsi}, as above] maps solutions to equation~\eqref{eqn:radial_teukolsky} for $s = -1$ to solutions for $s = 1$.
Similarly, $\pb{-1} M^a \overline{\pb{-1} \tau_a^\dagger}$ maps solutions $s = 1$ to $s = -1$.
In both cases, these maps preserve the value of $p$ of the solutions.
That is,

\begin{equation} \label{eqn:radial_tsi}
  \Delta \mathscr{D}_{0(\pm m)(\pm \omega)}^2 \Delta^{(1 \pm 1)/2} \pb{\pm 1} \widehat{\Omega}_{lm\omega p} (r) = \pb{\pm 1} C_{lm\omega p} \pb{\mp 1} \Delta^{(1 \mp 1)/2} \pb{\mp 1} \widehat{\Omega}_{lm\omega p}.
\end{equation}
There is a relationship between the constants $B_{lm\omega}$ and $\pb{\pm 1} C_{lm\omega p}$.
To see this, note that from equation~\eqref{eqn:conjugate_modes_parity},~\eqref{eqn:angular_tsi}, and~\eqref{eqn:radial_tsi},

\begin{subequations} \label{eqn:decoupled_action}
  \begin{align}
    \pb{\mp 1} M^a \overline{\pb{\pm 1} \tau_a^\dagger \pb{\mp 1} \Omega} &= -\frac{1}{4} \int_{-\infty}^\infty \ud \omega \sum_{l = 1}^\infty \sum_{|m| \leq l} \sum_{p = \pm 1} p B_{lm\omega} e^{i(m\phi - \omega t)} \pb{\mp 1} \Theta_{lm\omega} (\theta) \pb{\mp 1} \widehat{\Omega}_{lm\omega p} (r), \\
    \pb{\mp 1} M^a \overline{\pb{\mp 1} \tau_a^\dagger \pb{\pm 1} \Omega} &= -\frac{1}{2^{2 \pm 1}} \int_{-\infty}^\infty \ud \omega \sum_{l = 1}^\infty \sum_{|m| \leq l} \sum_{p = \pm 1} p \pb{\pm 1} C_{lm\omega p} e^{i(m\phi - \omega t)} \pb{\mp 1} \Theta_{lm\omega} (\theta) \pb{\mp 1} \widehat{\Omega}_{lm\omega p} (r).
  \end{align}
\end{subequations}
Now, equation~\eqref{eqn:tsr} implies that, for $|s| = 1$,

\begin{equation}
  \pb{s} M^a \overline{\pb{s} \tau_a^\dagger \pb{-s} \Omega} = \pb{s} M^a \overline{\pb{-s} \tau_a^\dagger \pb{s} \Omega},
\end{equation}
and so

\begin{equation} \label{eqn:C_from_B}
  \pb{\pm 1} C_{lm\omega p} = 2^{\pm 1} B_{lm\omega}.
\end{equation}
Thus, in particular, we find that

\begin{equation} \label{eqn:radial_tsi_no_p}
  \Delta \mathscr{D}_{0(\pm m)(\pm \omega)}^2 \Delta^{(1 \pm 1)/2} \pb{\pm 1} \widehat{\Omega}_{lm\omega} (r) = 2^{\pm 1} B_{lm\omega} \pb{\mp 1} \Delta^{(1 \mp 1)/2} \pb{\mp 1} \widehat{\Omega}_{lm\omega} (r).
\end{equation}

At this point, in equations~\eqref{eqn:decoupled_action} and~\eqref{eqn:C_from_B}, we have determined the action of the symmetry operators $\pb{s} M^a \overline{\pb{s} \tau_a^\dagger}$ and $\pb{s} M^a \overline{\pb{-s} \tau_a^\dagger}$ on the expansion of the (complex conjugated) master variables.
However, one can also show that $\pb{s} \boldsymbol{\mathcal C}$ is diagonal on the space of vector potentials, assuming a particular expansion of the vector potential.
We now construct this expansion, assuming (for simplicity) that the vector potential is purely radiative.

First, given a master variable $\pb{s} \Omega$ which comes from the vector potential, we need to define Debye potentials $\pb{\pm 1} \psi$ such that

\begin{equation}
  \pb{s} \Omega = \pb{s} M^a \overline{\pb{s} \tau_a^\dagger \pb{-s} \psi} = \pb{s} M^a \overline{\pb{-s} \tau_a^\dagger \pb{s} \psi}.
\end{equation}
This can be done by using an expansion for $\pb{s} \psi$ of the form~\eqref{eqn:decoupled_modes}, where

\begin{equation} \label{eqn:debye_def}
  \pb{s} \widehat{\psi}_{lm\omega p} (r) = -\frac{4}{p B_{lm\omega}} \pb{s} \widehat{\Omega}_{lm\omega p} (r),
\end{equation}
as a consequence of the linearity in each term in the expansions in equations~\eqref{eqn:decoupled_action}.
Define the vector potentials $\pb{\pm 1} \var A_a$ by

\begin{equation}
  \pb{\pm 1} \var A_a = \pb{\pm 1} \tau_a^\dagger \pb{\mp 1} \psi,
\end{equation}
so that there is now a natural expansion for $\pb{\pm 1} \var A_a$ of the following form:

\begin{equation} \label{eqn:A_modes}
  \pb{\pm 1} \var A_a = \int_{-\infty}^\infty \ud \omega \sum_{l = 1}^\infty \sum_{|m| \leq l} \sum_{p = \pm 1} (\pb{\pm 1} \var A_{lm\omega p})_a,
\end{equation}
where

\begin{equation}
  (\pb{\pm 1} \var A_{lm\omega p})_a = \pb{\pm 1} \tau_a^\dagger \left[e^{i(m\phi - \omega t)} \pb{\mp 1} \Theta_{lm\omega} (\theta) \pb{\mp 1} \widehat{\psi}_{lm\omega p} (r)\right].
\end{equation}
The vector potentials $\overline{\pb{\pm 1} \var A_a}$ are both vector potentials, like $\var A_a$, which yield $\pb{s} \Omega$ when acted upon by $\pb{s} M^a$ as in equation~\eqref{eqn:defM}, and so they are the same up to gauge.
The gauge conditions that these vector potentials satisfy are given by

\begin{equation}
  \overline{\pb{1} \var A_a} l^a = 0, \qquad \overline{\pb{-1} \var A_a} n^a = 0,
\end{equation}
which are (respectively) called the \emph{ingoing} and \emph{outgoing radiation gauge condition}.

We finally consider the action of our symmetry operators $\pb{s} \boldsymbol{\mathcal C}$ on $\overline{\pb{\pm 1} \var A_a}$.
We find that

\begin{equation} \label{eqn:C_action}
  \begin{split}
    \pb{s} \mathcal{C}_a{}^b \overline{\pb{\pm 1} \var A_a} &= \pb{s} \tau^\dagger_a \pb{-s} M^b \overline{\pb{\pm 1} \tau_b^\dagger \pb{\mp 1} \psi} = \pb{s} \tau^\dagger_a \pb{-s} \Omega \\
    &= -\frac{1}{4} \int_{-\infty}^\infty \ud \omega \sum_{l = 1}^\infty \sum_{|m| \leq l} \sum_{p = \pm 1} p B_{lm\omega} \pb{s} \tau^\dagger_a \left[e^{i(m\phi - \omega t)} \pb{s} \Theta_{lm\omega} (\theta) \pb{s} \widehat{\psi}_{lm\omega p} (r)\right] \\
    &= -\frac{1}{4} \int_{-\infty}^\infty \ud \omega \sum_{l = 1}^\infty \sum_{|m| \leq l} \sum_{p = \pm 1} p B_{lm\omega} (\pb{s} A_{lm\omega p})_a.
  \end{split}
\end{equation}
This shows that, apart from a complex conjugation, the effect of applying the symmetry operator $\pb{s} \boldsymbol{\mathcal C}$ is to multiply each term in this expansion by a constant.
We can add monopole terms to $\pb{\pm 1} \var A_a$ in equation~\eqref{eqn:C_action}, but this will not change the result since $\pb{\pm 1} \boldsymbol{\mathcal{C}}$ depends only on $\var \phi_0$ and $\var \phi_2$ (which only have $l \geq 1$).

\section{Conserved currents} \label{sec:currents}

\subsection{Previous currents}

In terms of symmetry operators, two types of conserved currents were constructed in~\cite{Andersson:2015xla}: for two independent variations $\var_1$ and $\var_2$, and for two symmetry operators $\pp{-} \boldsymbol{\mathcal{O}}$ and $\pp{+} \boldsymbol{\mathcal{O}}$ of the first and second kind, respectively, they showed that the currents defined by

\begin{subequations}
  \begin{align}
    \pb{\textrm{EZ}} j_{AA'} [\pp{-} \boldsymbol{\mathcal{O}}, \var_1 \boldsymbol{A}, \var_2 \boldsymbol{A}] &\equiv \var_1 \phi_{BA} \overline{\pp{-} \mathcal{O}^B{}_{A'}{}^c \var_2 A_c}, \label{eqn:andersson_current_EZ} \\
    \pb{\textrm{C}} j_{AA'} [\pp{+} \boldsymbol{\mathcal{O}}, \var_1 \boldsymbol{A}, \var_2 \boldsymbol{A}] &\equiv \var_1 \bar{\chi}_{B'A'} \overline{\pp{+} \mathcal{O}_A{}^{B' c} \var_2 A_c},
  \end{align}
\end{subequations}
were conserved, as a consequence of equation~\eqref{eqn:maxwell_spinor}, $\pb{1} \boldsymbol{\mathcal{F}} \cdot \pp{+} \boldsymbol{\mathcal{O}} = 0$, and $\overline{\pb{1} \boldsymbol{\mathcal{F}}} \cdot \pp{-} \boldsymbol{\mathcal{O}} = 0$.
The prefixes ``EZ'' and ``C'' for these currents denote the fact that these currents, in the classification scheme of~\cite{Andersson:2015xla} and Section~\ref{sec:geometric_optics}, are either ``energy/zilch'' currents or ``chiral'' currents.

Another natural means of generating conserved currents is to use a symmetry operator in conjunction with a conserved stress-energy tensor.
For example, we have that

\begin{equation}
  T_{ab} [\var_1 \boldsymbol{A}, \var_2 \boldsymbol{A}] = \frac{1}{2\pi} \var_1 \phi_{AB} \var_2 \bar{\phi}_{A'B'}
\end{equation}
is a conserved stress-energy tensor, satisfying $\nabla^a T_{ab} = 0$ for any pair of variations $\var_1$ and $\var_2$.
The factor of $2\pi$ is conventional, and present in order for it to reduce to the normal stress-energy tensor of electromagnetism when $\var_1 = \var_2$.
Given a Killing vector $\xi^a$, one can construct a conserved current

\begin{equation}
  \pb{\xi} j^a [\var_1 \boldsymbol{A}, \var_2 \boldsymbol{A}] = 2\pi \xi_b T^{ab} [\var_1 \boldsymbol{A}, \var_2 \boldsymbol{A}].
\end{equation}
The current in equation (5.11) of~\cite{Andersson:2015xla} is of this form.
Note also that

\begin{equation}
  \pb{\textrm{EZ}} j_a [\pb{\xi} \boldsymbol{L}, \var_1 \boldsymbol{A}, \var_2 \boldsymbol{A}] = \pb{\xi} j_a [\var_1 \boldsymbol{A}, \var_2 \boldsymbol{A}].
\end{equation}

\subsection{Symplectic products} \label{sec:symplectic}

Given a theory that possesses a Lagrangian formulation, one method of generating conserved quantities is to use the symplectic product.
Following Burnett and Wald~\cite{Burnett57}, we start with a general Lagrangian four-form $\boldsymbol{L} [\boldsymbol{\psi}, \boldsymbol{\phi}] = \dual \mathcal{L}$ that is locally constructed from background fields $\boldsymbol{\psi}$ and dynamical fields $\boldsymbol{\phi}$.
We consider a variation defined by $\var$, and we suppose that the variation of the Lagrangian obeys

\begin{equation}
  \var \boldsymbol{L} [\boldsymbol{\psi}, \boldsymbol{\phi}] \equiv \boldsymbol{E} [\boldsymbol{\psi}, \boldsymbol{\phi}] \cdot \var \boldsymbol{\phi} + \ud \boldsymbol{\Theta} [\boldsymbol{\psi}, \boldsymbol{\phi}, \var \boldsymbol{\phi}],
\end{equation}
where $\boldsymbol{\Theta}$ is the \emph{symplectic potential}, and $\boldsymbol{E}$ is a tensor-valued four-form, that is, it is a tensor with four antisymmetric indices and a number of indices that matches that of $\var \boldsymbol{\phi}$, such that $\boldsymbol{E} \cdot \var \boldsymbol{\phi}$ is a four-form.
On shell, $\boldsymbol{E} = 0$, which makes the integral of $\var \boldsymbol{L}$ a boundary term.
Given two variations $\var_1$ and $\var_2$ (which must commute), define the \emph{symplectic current} by

\begin{equation}
  \boldsymbol{\omega} [\boldsymbol{\psi}, \boldsymbol{\phi}, \var_1 \boldsymbol{\phi}, \var_2 \boldsymbol{\phi}] \equiv \var_1 \boldsymbol{\Theta} [\boldsymbol{\psi}, \boldsymbol{\phi}, \var_2 \boldsymbol{\phi}] - \var_2 \boldsymbol{\Theta} [\boldsymbol{L}, \boldsymbol{\psi}, \boldsymbol{\phi}, \var_1 \boldsymbol{\phi}].
\end{equation}
Thus, since $\ud$ and $\var$ commute,

\begin{equation}
  \ud \boldsymbol{\omega} = \var_1 (\var_2 \boldsymbol{L} - \boldsymbol{E} \cdot \var_2 \boldsymbol{\phi}) - \var_2 (\var_1 \boldsymbol{L} - \boldsymbol{E} \cdot \var_1 \boldsymbol{\phi}) = \var_2 \boldsymbol{E} \cdot \var_1 \boldsymbol{\phi} - \var_1 \boldsymbol{E} \cdot \var_2 \boldsymbol{\phi}.
\end{equation}
However, if $\var_1 \boldsymbol{\phi}$ and $\var_2 \boldsymbol{\phi}$ are both solutions to the linearized equations, so that $\var_1 \boldsymbol{E} = \var_2 \boldsymbol{E} = 0$, then $\ud \boldsymbol{\omega} = 0$; that is, $\boldsymbol{\omega}$ is a conserved current.

In the case of the electromagnetic field, we have that

\begin{equation}
  \boldsymbol{L}_{\textrm{M}} = \frac{1}{4} F_{ab} F^{ab} \boldsymbol{\epsilon} = \frac{1}{2} \boldsymbol{F} \wedge \dual \boldsymbol{F},
\end{equation}
and so

\begin{equation}
  \var \boldsymbol{L}_{\textrm{M}} = \ud \var \boldsymbol{A} \wedge \dual \boldsymbol{F} = \var \boldsymbol{A} \wedge \ud \dual \boldsymbol{F} + \ud (\var \boldsymbol{A} \wedge \dual \boldsymbol{F}).
\end{equation}
Thus, we find that the symplectic current, which we denote by $\boldsymbol{\omega}_{\textrm M}$, is given by

\begin{equation}
  \boldsymbol{\omega}_{\textrm M} (\var_1 \boldsymbol{A}, \var_2 \boldsymbol{A}) = \var_1 \boldsymbol{A} \wedge \dual \var_2 \boldsymbol{F} - \var_2 \boldsymbol{A} \wedge \dual \var_1 \boldsymbol{F}.
\end{equation}
We define the corresponding vector current by

\begin{equation} \label{eqn:symplectic_current}
  \pb{S} j^a \left[\var_1 \phi, \var_2 \phi\right] \equiv \left(\dual \boldsymbol{\omega} \left[\var_1 \phi, \var_2 \phi\right]\right)^a.
\end{equation}

Note that this quantity is not gauge-invariant, in the sense that it is not invariant under a change $\var_1 \boldsymbol{A} \to \var_1 \boldsymbol{A} + \ud \lambda$.
However, we have that, defining $\var_1 = \var$ and $\var_2 \boldsymbol{A} = \ud \lambda$,

\begin{equation}
  \boldsymbol{\omega}_{\textrm M} (\var \boldsymbol{A}, \ud \lambda) = -\ud \lambda \wedge \dual \var \boldsymbol{F} = -\ud (\lambda \dual \var \boldsymbol{F}),
\end{equation}
where we have used that $\ud \dual \var \boldsymbol{F} = 0$ on shell, and so $\boldsymbol{\omega}$ is gauge-invariant up to a boundary term, a result that carries over to linearized gravity.

We further note that the symplectic current is equivalent to the conserved currents discussed in~\cite{Andersson:2015xla}, up to a boundary term.
Let $\var_2 \boldsymbol{A} = \overline{\boldsymbol{\mathcal{O}} \cdot \var \boldsymbol{A}} \equiv \overline{\pp{+} \boldsymbol{\mathcal{O}} \cdot \var \boldsymbol{A}} + \overline{\pp{-} \boldsymbol{\mathcal{O}} \cdot \var \boldsymbol{A}}$ for two symmetry operators, $\pp{+} \boldsymbol{\mathcal{O}}$ and $\pp{-} \boldsymbol{\mathcal{O}}$, where

\begin{equation}
  \dual \ud \pp{\pm} \boldsymbol{\mathcal{O}} = \pm i \ud \pp{\pm} \boldsymbol{\mathcal{O}}
\end{equation}
that is, we break up $\var_2 F$ into its self-dual and anti-self-dual parts.
Then we have that

\begin{equation}
  \begin{split}
    \boldsymbol{\omega}_{\textrm M} \left[\var \boldsymbol{A}, \overline{\boldsymbol{\mathcal{O}} \cdot \var \boldsymbol{A}}\right] &= i \var \boldsymbol{A} \wedge \ud \left(\overline{\pp{-} \boldsymbol{\mathcal{O}} \cdot \var \boldsymbol{A}} - \overline{\pp{+} \boldsymbol{\mathcal{O}} \cdot \var \boldsymbol{A}}\right) - \overline{\boldsymbol{\mathcal{O}} \cdot \var \boldsymbol{A}} \wedge \dual \var \boldsymbol{F} \\
    &= \left(i \var \boldsymbol{F} - \dual \var \boldsymbol{F}\right) \wedge \overline{\pp{-} \boldsymbol{\mathcal{O}} \cdot \var \boldsymbol{A}} - \left(i \var \boldsymbol{F} + \dual \var \boldsymbol{F}\right) \wedge \overline{\pp{+} \boldsymbol{\mathcal{O}} \cdot \var \boldsymbol{A}} + \ud \boldsymbol{\alpha},
  \end{split}
\end{equation}
where

\begin{equation}
  \boldsymbol{\alpha} \equiv i \var \boldsymbol{A} \wedge \left(\overline{\pp{-} \boldsymbol{\mathcal{O}} \cdot \var \boldsymbol{A}} - \overline{\pp{+} \boldsymbol{\mathcal{O}} \cdot \var \boldsymbol{A}}\right).
\end{equation}
Note that the terms in parentheses are related to the self-dual and anti-self-dual parts of $\var \boldsymbol{F}$:

\begin{equation}
  i \var \boldsymbol{F} \pm \dual \var \boldsymbol{F} = -2i \pp{\pm} [\var \boldsymbol{F}] = \mp 2 \dual \pp{\pm} [\var \boldsymbol{F}],
\end{equation}
so that

\begin{equation}
  \boldsymbol{\omega}_{\textrm M} \left[\var \boldsymbol{A}, \overline{\boldsymbol{\mathcal{O}} \cdot \var \boldsymbol{A}}\right] = 2\left[\left(\dual \pp{+} [\var \boldsymbol{F}]\right) \wedge \overline{\pp{-} \boldsymbol{\mathcal{O}} \cdot \var \boldsymbol{A}} + \left(\dual \pp{-} [\var \boldsymbol{F}]\right) \wedge \overline{\pp{+} \boldsymbol{\mathcal{O}} \cdot \var \boldsymbol{A}}\right] + \ud \boldsymbol{\alpha}.
\end{equation}
Moreover, we have that

\begin{equation}
  [\dual (\var \boldsymbol{A} \wedge \dual \var \boldsymbol{F})]_a = -\frac{1}{2} \var A^b \var F_{ab} = \frac{1}{2} \left(\var A_A{}^{B'} \var \bar{\chi}_{A'B'} + \var A^B{}_{A'} \var \phi_{AB}\right),
\end{equation}
and so one finds that

\begin{equation} \label{eqn:andersson_symp}
    \pb{S} j_a \left[\boldsymbol{L}_{\textrm{M}}, \var \boldsymbol{A}, \overline{\boldsymbol{\mathcal{O}} \cdot \var \boldsymbol{A}}\right] = \pb{\textrm{EZ}} j_a [\pp{-} \boldsymbol{\mathcal{O}}, \var \boldsymbol{A}, \var \boldsymbol{A}] - \pb{\textrm{C}} j_a [\pp{+} \boldsymbol{\mathcal{O}}, \var \boldsymbol{A}, \var \boldsymbol{A}] + (\dual \ud \boldsymbol{\alpha})_a.
\end{equation}
Thus, up to a non-gauge-invariant boundary term $\ud \boldsymbol{\alpha}$, the symplectic current has the same form as that given by~\cite{Andersson:2015xla}.

\subsection{Summary of currents} \label{sec:new_currents}

We now list the currents defined in this paper.
Our new current is given by

\begin{equation} \label{eqn:pm_1_j}
  \pb{\pm 1} j^a [\var \boldsymbol{A}] \equiv \pb{S} j^a \left[\pb{\pm 1} \boldsymbol{\mathcal C} \cdot \var \boldsymbol{A}, \overline{\pb{\pm 1} \boldsymbol{\mathcal C} \cdot \var \boldsymbol{A}}\right],
\end{equation}
where $\pb{\pm 1} \boldsymbol{\mathcal C}$ is defined in equation~\eqref{eqn:C_def}, and the symplectic product in equation~\eqref{eqn:symplectic_current}.
The particular structure of this current is such that it yields a simple result in the limit of geometric optics.
Although it is not new, we will also consider the following current given in~\cite{Andersson:2015xla}, which is defined by

\begin{equation} \label{eqn:A_j}
  \pb{\mathcal A} j^a [\var \boldsymbol{A}] \equiv \pb{\textrm{EZ}} j^a [\boldsymbol{\mathcal A}, \var \boldsymbol{A}, \var \boldsymbol{A}],
\end{equation}
where $\boldsymbol{\mathcal A}$ is the symmetry operator defined in equation~\eqref{eqn:def_A} and $\pb{\textrm{EZ}} j^a$ is defined by equation~\eqref{eqn:andersson_current_EZ}.

The key property of these conserved currents are that they are related to the Carter constant in the geometric optics limit that we will discuss in Section~\ref{sec:geometric_optics}.
Explicitly, we define the charge $\pb{\cdots} Q [\Sigma, \var \boldsymbol{A}]$ for some current $\pb{\cdots} j^a [\Sigma, \var \boldsymbol{A}]$ and some spacelike hypersurface $\Sigma$:

\begin{equation} \label{eqn:charge_def}
  \pb{\cdots} Q[\Sigma, \var \boldsymbol{A}] \equiv \int_\Sigma \pb{\cdots} j^a [\var \boldsymbol{A}] \ud^3 \Sigma_a.
\end{equation}
In the geometric optics limit, these charges are all related to the sum of either the Carter constants (for $\pb{\mathcal A} Q[\Sigma, \var \boldsymbol{A}]$) or the squares of the Carter constants (for $\pb{\pm 1} Q[\Sigma, \var \boldsymbol{A}]$) of photons passing through $\Sigma$.

Another property of these currents that we consider is their integrals over portions of null infinity and the horizon, or \emph{fluxes}.
Our results in Section~\ref{sec:fluxes} are that these fluxes are finite for $\pb{\pm 1} j^a [\var \boldsymbol{A}]$, but infinite for $\pb{\mathcal A} j^a [\var \boldsymbol{A}]$.

\section{Geometric Optics} \label{sec:geometric_optics}

\subsection{Formalism}

We now review the geometric optics (or high-frequency/eikonal) approximation for source-free solutions to Maxwell's equations.
We assume a one-parameter family of complex, Lorenz gauge solutions of the form

\begin{equation}
  \var A_a = [a \varpi_a + O(\epsilon)] e^{-i \vartheta/\epsilon},
\end{equation}
where $\varpi_a$ is constrained by the normalization condition $\varpi_a \bar{\varpi}^a = -1$ and $a$ is real.
Inserting this ansatz into the Lorenz gauge condition and Maxwell's equations and equating coefficients of powers of $\epsilon$ we find (see, for example, Misner, Thorne, and Wheeler~\cite{mtw}):

\begin{itemize}

\item[i.] The vector $k^a$ defined by

  \begin{equation}
    k_a \equiv \nabla_a \vartheta,
  \end{equation}
  which represents the wavevector for an electromagnetic wave, describes a congruence of null geodesics:

  \begin{equation}
    k_a k^a = 0, \qquad k^b \nabla_b k^a = 0.
  \end{equation}

\item[ii.] The polarization vector $\varpi^a$ is orthogonal to $k^a$ and is parallel-transported along these geodesics:

  \begin{equation} \label{eqn:polarization_conditions}
    k^a \varpi_a = 0, \qquad k^b \nabla_b \varpi^a = 0.
  \end{equation}

\item[iii.] The amplitude $a$ evolves according to

  \begin{equation} \label{eqn:etendue}
    \nabla_a (a^2 k^a) = 0.
  \end{equation}

\end{itemize}

Next, since $k^a$ is null, we can write it as

\begin{equation}
  k^{AA'} = \kappa^A \bar{\kappa}^{A'},
\end{equation}
for some spinor $\kappa^A$.
We choose a second spinor $\lambda^A$ so that $(\kappa, \lambda)$ form a spin basis like $(o, \iota)$.
We can write the polarization vector $\varpi^a$ on this basis as

\begin{equation}
  \varpi^{AA'} = \alpha \kappa^A \bar{\kappa}^{A'} + \beta \lambda^A \bar{\lambda}^{A'} + e_R \kappa^A \bar{\lambda}^{A'} + e_L \lambda^A \bar{\kappa}^{A'},
\end{equation}
for some complex coefficients $\alpha$, $\beta$, $e_R$, and $e_L$.
From the condition in equation~\eqref{eqn:polarization_conditions}, $\beta = 0$, and similarly we can set $\alpha = 0$ by the gauge transformation $\var A_a \to \var A_a + \nabla_a \lambda$, where $\lambda = -ia \alpha \epsilon e^{-i\vartheta/\epsilon}$, which maintains the Lorenz gauge condition.

The coefficients $e_L$ and $e_R$ parameterize the left and right circularly polarized components of the radiation.
To see this, form a tetrad from $\kappa_A$ and $\lambda_A$ in the form

\begin{equation}
  \begin{aligned}
    t_{AA'} = \frac{1}{\sqrt{2}} (\kappa_A \bar{\kappa}_{A'} + \lambda_A \bar{\lambda}_{A'}), &\qquad z_{AA'} = \frac{1}{\sqrt{2}} (\kappa_A \bar{\kappa}_{A'} - \lambda_A \bar{\lambda}_{A'}), \\
    x_{AA'} = \frac{1}{\sqrt{2}} (\kappa_A \bar{\lambda}_{A'} + \lambda_A \bar{\kappa}_{A'}), &\qquad y_{AA'} = \frac{i}{\sqrt{2}} (\kappa_A \bar{\lambda}_{A'} - \lambda_A \bar{\kappa}_{A'}).
  \end{aligned}
\end{equation}
It then follows that

\begin{equation}
  \varpi_a = \frac{e_R}{\sqrt{2}} (x_a - i y_a) + \frac{e_L}{\sqrt{2}} (x_a + i y_a).
\end{equation}
That is, on this basis, $e_R = 1$, $e_L = 0$ corresponds to right circularly polarized light and $e_L = 1$, $e_R = 0$ corresponds to left circularly polarized light.

We next compute the spinor fields $\var \phi_{AB}$ and $\var \bar{\chi}_{A'B'}$ that are defined by equations~\eqref{eqn:faraday_spinor}.
To leading order in $\epsilon$ we find that

\begin{equation} \label{eqn:complex_geometric_optics}
  \var \phi_{AB} = \frac{1}{\epsilon} ia e_R \kappa_A \kappa_B e^{-i\vartheta/\epsilon} + O(1), \qquad \var \bar{\chi}_{A'B'} = \frac{1}{\epsilon} ia e_L \bar{\kappa}_A \bar{\kappa}_B e^{-i\vartheta/\epsilon} + O(1),
\end{equation}
so that each spinor field contains only one circular polarization.
In the case of real fields, one is only concerned with $\var \phi_{AB}$, and one finds that

\begin{equation} \label{eqn:real_geometric_optics}
  \var \phi_{AB} = \frac{1}{\epsilon} ia \kappa_A \kappa_B \left(e_R e^{-i\vartheta/\epsilon} - \bar{e}_L e^{i\vartheta/\epsilon}\right) + O(1).
\end{equation}
Note that equations~\eqref{eqn:complex_geometric_optics} and~\eqref{eqn:real_geometric_optics} are invariant under the rotation $\kappa_A \to e^{i\varphi} \kappa_A$, $\lambda_A \to e^{-i\varphi} \lambda_A$, since $e_R \to e^{-2i\varphi} e_R$ and $e_L \to e^{2i\varphi} e_L$.
We will be considering only real vector potentials, since only those are physical, although we have constructed complex vector potentials via symmetry operators.

\subsection{Conserved currents in geometric optics}

When we evaluate nonlinear quantities in the geometric optics limit, such as currents (which are quadratic), we discard all rapidly oscillating terms, in effect taking a spacetime average over a scale large compared to $\epsilon$ but small compared to the curvature scale of the background.\footnote{See~\cite{Burnett:1989gp} for a more rigorous treatment of this averaging procedure involving weak limits.}
Such averages will be denoted by $\langle j^a\rangle$, for currents $j^a$.

A simple example, for a Killing vector $\xi^a$, is given by $\xi_b T^{ab} [\var \boldsymbol{A}, \var \boldsymbol{A}]$.
In the limit of geometric optics this current becomes

\begin{equation} \label{eqn:energy_example}
  \begin{split}
    \langle \xi_b T^{ab} [\var \boldsymbol{A}, \var \boldsymbol{A}]\rangle &= \frac{1}{\epsilon^2} \frac{a^2}{2\pi} k^a k^b \xi_b (|e_R|^2 + |e_L|^2) + O(\epsilon^{-1}) = \frac{1}{\epsilon^2} \frac{E_\xi}{2\pi \hbar} a^2 k^a + O(\epsilon^{-1}),
  \end{split}
\end{equation}
where $E_\xi \equiv \xi_a p^a$ is the conserved quantity with respect to $\xi_a$ of a given photon with wavevector $k^a = p^a/\hbar$.
Although this calculation is purely classical, a factor of $\hbar$ occurs in converting between expressions involving $k^a$ and conserved quantities (like $E_\xi$) defined using the four-momentum $p^a$.
Considering the electromagnetic field in this limit as a null fluid of photons with number density $\mathcal N^a$, one finds that~\cite{mtw}

\begin{equation}
  \frac{1}{\epsilon^2} a^2 k^a = 2\pi \hbar \mathcal N^a,
\end{equation}
which implies that

\begin{equation}
  \langle \xi_b T^{ab} [\var \boldsymbol{A}, \var \boldsymbol{A}]\rangle = E_\xi \mathcal N^a + O(\epsilon^{-1}).
\end{equation}
This gives the expected result that integrating the current~\eqref{eqn:energy_example} over a hypersurface gives the total $E_\xi$ (say, energy, in the case $\vec{\xi} = \partial_t$) of the photons crossing the hypersurface.

This example has two interesting properties.
First, this is an example of a conserved current $j^a$ that reduces in geometric optics to

\begin{equation} \label{eqn:conserved_quantity}
  \langle j^a\rangle = \frac{1}{\epsilon^n} Q a^2 k^a + O(\epsilon^{-n + 1}),
\end{equation}
for some quantity $Q$.
Conversely, one can show, from equation~\eqref{eqn:etendue}, that if $\nabla_a j^a = 0$ and equation~\eqref{eqn:conserved_quantity} holds, then $Q$ is a conserved quantity along the integral curves of $k^a$.
The condition~\eqref{eqn:conserved_quantity} is satisfied by all currents that we consider in this paper.

Another property of this example is that the current~\eqref{eqn:energy_example} is not dependent on the circular polarization parameters $e_R$ and $e_L$.
We will classify quadratic currents in the geometric optics limit in terms of their dependence on these parameters: currents that are independent will be called \emph{energy currents}, currents that are proportional to $|e_R|^2 - |e_L|^2$ will be called \emph{zilch currents}, and currents that are proportional to $e_R \bar{e}_L$ or $e_L \bar{e}_R$ will be called \emph{chiral} or \emph{anti-chiral  currents}, respectively.
The simplest example of a zilch current is the current

\begin{equation} \label{eqn:zilch_ex}
  \begin{split}
    \langle\xi_b T^{ab} [\var \boldsymbol{A}, \pb{\xi} \boldsymbol{L} \cdot \var \boldsymbol{A}]\rangle = \frac{1}{\epsilon^3} \frac{E_\xi^2}{2\pi \hbar^2} (|e_R|^2 - |e_L|^2) a^2 k^a + O(\epsilon^{-2}),
  \end{split}
\end{equation}
where the symmetry operator $\pb{\xi} \boldsymbol{L}$ is defined in equation~\eqref{eqn:def_xiL}.

Equation~\eqref{eqn:zilch_ex} also shows that zilch currents can yield conserved quantities in geometric optics that are quadratic in the four-momentum.
One can show this generally: energy currents yield conserved quantities that have odd powers of the $k^a$, zilch currents yield conserved quantities that have even powers of the $k^a$, and chiral and antichiral currents yield conserved quantities that depend on $\kappa^A$ and $\bar{\kappa}^{A'}$ individually, in addition to $k^a$.
As such, all of the currents that we consider in this paper are zilch currents.

We now consider the geometric optics limits of the currents that we defined in Section~\ref{sec:new_currents}.
We begin with the conserved current $\pb{\mathcal A} j^a [\var \boldsymbol{A}]$ in equation~\eqref{eqn:A_j}.
Using equations~\eqref{eqn:def_A} and~\eqref{eqn:andersson_current_EZ}, one finds that

\begin{equation} \label{eqn:A_j_optics}
  \langle\pb{\mathcal A} j^a [\var \boldsymbol{A}]\rangle = -\frac{1}{\epsilon^3} \frac{iK}{4\hbar^2} (|e_R|^2 - |e_L|^2) a^2 k^a + O(\epsilon^{-2}).
\end{equation}
This moreover uses the fact that

\begin{equation}
  \left|\zeta_{AB} \kappa^A \kappa^B\right|^2 = \frac{1}{2} K/\hbar^2.
\end{equation}
Consider now the charge $\pb{\mathcal A} Q[\Sigma, \var \boldsymbol{A}]$ obtained by integrating this current over a hypersurface $\Sigma$.
As in the case of the example~\eqref{eqn:energy_example}, equation~\eqref{eqn:A_j_optics} means that $\pb{\mathcal A} Q[\Sigma, \var \boldsymbol{A}]$ is proportional to the sum of the Carter constants of the photons passing through $\Sigma$.
Considering the case of the current $\pb{\pm 1} j^a [\var \boldsymbol{A}]$ in equation~\eqref{eqn:pm_1_j}, we note that

\begin{equation}
  \pb{1} \mathcal{C}^{AA'}{}_b \var A^b = -\frac{1}{\epsilon^2} \frac{a}{2} \zeta^2 o^A o^B \kappa_B \bar{\kappa}^{A'} (\iota_C \kappa^C)^2 \left(e_R e^{-i\vartheta/\epsilon} + \bar{e}_L e^{i\vartheta/\epsilon}\right),
\end{equation}
and so

\begin{equation} \label{eqn:pm_1_j_optics}
  \langle\pb{1} j_a [\var \boldsymbol{A}]\rangle = \frac{1}{\epsilon^5} \frac{iK^2}{16\hbar^4} (|e_R|^2 - |e_L|^2) a^2 k_a + O(\epsilon^{-4}).
\end{equation}
The interpretation of this expression is that the charge $\pb{1} Q[\Sigma, \var \boldsymbol{A}]$ is proportional to the sum of the squares of the Carter constants of the photons passing through $\Sigma$.
Moreover, one can easily show that, at least in the limit of geometric optics,

\begin{equation}
  \langle\pb{-1} j_a [\var \boldsymbol{A}]\rangle = \langle\pb{1} j_a [\var \boldsymbol{A}]\rangle + O(\epsilon^{-4}),
\end{equation}
and so we do not need to separately compute the current $\pb{-1} j_a [\var \boldsymbol{A}]$.

\section{Fluxes at null infinity and the horizon} \label{sec:fluxes}

In order for a conserved current in Kerr to be physically useful, its fluxes should be finite when evaluated at null infinity and at the horizon.
In this section, we determine which of the currents in Section~\ref{sec:new_currents} satisfy this requirement.
As one might be interested in the particular values of these fluxes (if they are useful for calculating the rate of change of the Carter constant of a charged particle, for example), we also give the values of these fluxes in equations~\eqref{eqn:fluxes_pm_1_Q} (for $\pb{\pm 1} j^a [\overline{\pb{1} \var \boldsymbol{A}}, \Sigma]$), using the expansion for $\pb{1} \var A_a$ given in equation~\eqref{eqn:A_modes}.

\subsection{Definitions} \label{sec:flux_defs}

We define the fluxes of our currents through the horizon and null infinity by the following expressions (for details on how one arrives at these expressions, see appendix~A of~\cite{Grant2020}):

\begin{subequations} \label{eqn:volume_elements}
  \begin{align}
    \left.\frac{\ud^2 \pb{\cdots} Q}{\ud v \ud \Omega}\right|_{H^+} &\equiv \lim_{r \to r^+ \textrm{, fixed } v} \left(\Sigma n^a - \frac{1}{2} \Delta l^a\right) \pb{\cdots} j_a, \qquad &\left.\frac{\ud^2 \pb{\cdots} Q}{\ud u \ud \Omega}\right|_{H^-} &= \lim_{r \to r^+ \textrm{, fixed } u} \left(\Sigma n^a - \frac{1}{2} \Delta l^a\right) \pb{\cdots} j_a, \\
    \left.\frac{\ud^2 \pb{\cdots} Q}{\ud u \ud \Omega}\right|_{\scri^+} &= \lim_{r \to \infty \textrm{, fixed } u} r^2 \left(n^a - \frac{1}{2} l^a\right) \pb{\cdots} j_a, \qquad &\left.\frac{\ud^2 \pb{\cdots} Q}{\ud v \ud \Omega}\right|_{\scri^-} &= \lim_{r \to \infty \textrm{, fixed } v} r^2 \left(n^a - \frac{1}{2} l^a\right) \pb{\cdots} j_a.
  \end{align}
\end{subequations}
Here, the vector $\Sigma^a - \Delta l^a/2$ is equal to the volume element $\ud \Sigma^a$ for each of these surfaces, divided by either $\ud u \ud \Omega$ or $\ud v \ud \Omega$, where $\ud \Omega$ is the element of solid angle:

\begin{equation}
  \ud \Omega \equiv \sin \theta \ud \theta \begin{cases}
    \ud \psi & \textrm{ at} H^+, \scri^- \\
    \ud \chi & \textrm{ at} H^-, \scri^+
  \end{cases},
\end{equation}
where $\psi$ and $\chi$ are defined by

\begin{equation}
  \psi \equiv \phi + \int \frac{a \ud r}{\Delta}, \qquad \chi = \phi - \int \frac{a \ud r}{\Delta}.
\end{equation}

From equation~\eqref{eqn:volume_elements}, it is apparent that the relevant components of the currents constructed in this paper are those along $l_a$ and $n_a$.
In particular,

\begin{subequations} \label{eqn:pm_1_components}
  \begin{align}
    \pb{1} j_l \left[\var \boldsymbol{A}\right] = -i\Im[(\pb{1} \boldsymbol{\mathcal C} \cdot \var \boldsymbol{A})_{\bar m} (\overline{\pb{1} \boldsymbol{\mathcal F}} \cdot \pb{1} \boldsymbol{\mathcal C} \cdot \var \boldsymbol{A})_0], &\quad \pb{1} j_n \left[\var \boldsymbol{A}\right] = -i\Im[(\pb{1} \boldsymbol{\mathcal C} \cdot \var \boldsymbol{A})_n (\overline{\pb{1} \boldsymbol{\mathcal F}} \cdot \pb{1} \boldsymbol{\mathcal C} \cdot \var \boldsymbol{A})_1], \\
    \pb{-1} j_n \left[\var \boldsymbol{A}\right] = i\Im[(\pb{-1} \boldsymbol{\mathcal C} \cdot \var \boldsymbol{A})_m (\overline{\pb{1} \boldsymbol{\mathcal F}} \cdot \pb{-1} \boldsymbol{\mathcal C} \cdot \var \boldsymbol{A})_2], &\quad \pb{-1} j_l \left[\var \boldsymbol{A}\right] = i\Im[(\pb{-1} \boldsymbol{\mathcal C} \cdot \var \boldsymbol{A})_l (\overline{\pb{1} \boldsymbol{\mathcal F}} \cdot \pb{-1} \boldsymbol{\mathcal C} \cdot \var \boldsymbol{A})_1],
  \end{align}
\end{subequations}
as well as

\begin{subequations} \label{eqn:A_components}
  \begin{align}
    \pb{\mathcal A} j_l \left[\var \boldsymbol{A}\right] &= -\frac{\Sigma}{2} \left\{\var \bar{\phi}_0 [(\npDelta - 2\gamma + \mu - \bar{\mu}) \var \phi_0 + \rho \var \phi_2] + \var \bar{\phi}_1 \left(\delta - 2\beta + \pi - \tau + \frac{1}{3} \bar{\tau}\right) \var \phi_0\right\}, \\
    \pb{\mathcal A} j_n \left[\var \boldsymbol{A}\right] &= -\frac{\Sigma}{2} \left\{\var \bar{\phi}_2 [(D + 2\epsilon - \rho + \bar{\rho}) \var \phi_2 - \mu \var \phi_0] + \var \bar{\phi}_1 \left(\bar{\delta} + 2\alpha - \tau + \pi - \frac{1}{3} \bar{\pi}\right) \var \phi_2\right\}.
  \end{align}
\end{subequations}

\subsection{Computations} \label{sec:flux_comps}

Using the results Section~\ref{sec:flux_defs}, we now determine whether the fluxes of the currents defined in Section~\ref{sec:new_currents} are finite, and if so, we determine the values of these fluxes.

We start with determining which fluxes are finite.
For the current $\pb{\mathcal A} j^a [\var \boldsymbol{A}]$, the flux diverges by the following argument: the peeling theorem~\cite{penrose1988spinors} implies that there exist solutions such that $\var \phi_2 \sim 1/r$.
Since $\rho - \bar{\rho} \sim 1/r^2$,

\begin{equation}
  \Sigma \var \bar{\phi}_2 (D + 2\epsilon - \rho + \bar{\rho}) \var \phi_2 \sim 1/r
\end{equation}
at least, since $D$ can lower at most by one factor of $r$.
Equations~\eqref{eqn:A_components} and~\eqref{eqn:volume_elements}, imply that the flux through $\scri^+$ is dominated by this piece, which, when multiplied by $r^2$, diverges as $r$ in the limit $r \to \infty$.
The fluxes of the current $\pb{\pm 1} j^a [\overline{\pb{1} \var \boldsymbol{A}}]$ are finite.
Showing this can be most easily done using equations~\eqref{eqn:pm_1_components}, along with the falloffs of $\pb{\pm 1} \var A_a$ and the falloffs of

\begin{equation} \label{eqn:1_bar_chi}
  \pb{1} \var \bar{\chi}_i \equiv (\overline{\pb{1} \mathcal{F}} \cdot \pb{1} \var \boldsymbol{A})_i.
\end{equation}
These falloffs are given in Table~\ref{table:em_asymp}, which is constructed using the methods of Appendix~\ref{appendix:asymptotic}.

For currents $\pb{\cdots} j^a$ whose fluxes are finite, we use the following expansion of these fluxes:

\begin{subequations}
  \begin{align}
    \left\langle\left.\frac{\ud^2 \pb{\cdots} Q}{\ud u \ud \Omega}\right|_{\scri^+, H^-} \right\rangle_{u, \chi} &\equiv \int_{-\infty}^\infty \ud \omega \sum_{l, l' = 1}^\infty \sum_{|m| \leq l, l'} \sum_{p, p' = \pm 1} \left.\frac{\ud^2 \pb{\cdots} Q_{ll'm\omega pp'}}{\ud u \ud \Omega}\right|_{\scri^+, H^-}, \\
    \left\langle\left.\frac{\ud^2 \pb{\cdots} Q}{\ud v \ud \Omega}\right|_{\scri^-, H^+} \right\rangle_{v, \psi} &\equiv \int_{-\infty}^\infty \ud \omega \sum_{l, l' = 1}^\infty \sum_{|m| \leq l, l'} \sum_{p, p' = \pm 1} \left.\frac{\ud^2 \pb{\cdots} Q_{ll'm\omega pp'}}{\ud v \ud \Omega}\right|_{\scri^-, H^+},
  \end{align}
\end{subequations}
where on the left-hand side we average in either $u$ and $\psi$ or $v$ and $\chi$.
Similarly, we write the asymptotic form of the Debye potentials using the asymptotic form of solutions of the radial Teukolsky equation~\eqref{eqn:teukolsky}, yielding~\cite{1974ApJ...193..443T}

\begin{equation} \label{eqn:debye_expansion}
  \pb{s} \widehat{\psi}_{lm\omega p} (r) = \begin{cases}
    Z_{lm\omega p}^{\textrm{down}} e^{-i\omega r^*}/r + Z_{lm\omega p}^{\textrm{up}} e^{i\omega r^*}/r^{2s + 1} & r^* \to \infty \vspace{0.5em} \\
    Z_{lm\omega p}^{\textrm{in}} e^{-ik_{m\omega} r^*}/\Delta^s + Z_{lm\omega p}^{\textrm{out}} e^{ik_{m\omega} r^*} & r^* \to -\infty
  \end{cases},
\end{equation}
where $k_{m\omega} = \omega - am/(2Mr_+)$.
Similarly, consider the quantities $\pb{1} \var A_n$, $\pb{1} \var A_{\bar m}$, $\pb{-1} \var A_l$, $\pb{-1} \var A_m$, and $\pb{1} \var \bar{\chi}_1$, which (in general) we denote by $q$.
These quantities are all constructed linearly from a Debye potential $\pb{s} \psi$, and so possess an expansion as in equation~\eqref{eqn:A_modes}.
Expanding asymptotically, we define the asymptotic angular dependences $\pb{q} S^{\textrm{in}/\textrm{out}/\textrm{up}/\textrm{down}}_{lm\omega p} (\theta)$ by

\begin{equation}
  q(t, r, \theta, \phi) = \int_{-\infty}^\infty \ud \omega \sum_{l = 1}^\infty \sum_{|m| \leq l} \sum_{p, p' = \pm 1} e^{i(m\phi - \omega t)} \begin{cases}
    \begin{aligned}
      &Z_{lm\omega p}^{\textrm{down}} \pb{q} S^{\textrm{down}}_{lm\omega p} (\theta) r^{n_q^{\textrm{down}}} e^{-i\omega r^*} \\
      &\hspace{3em}+ Z_{lm\omega p}^{\textrm{up}} \pb{q} S^{\textrm{up}}_{lm\omega p} (\theta) r^{n_q^{\textrm{up}}} e^{i\omega r^*}
    \end{aligned} & r^* \to \infty \vspace{0.5em} \\
    \begin{aligned}
      &Z_{lm\omega p}^{\textrm{in}} \pb{q} S^{\textrm{in}}_{lm\omega p} (\theta) \Delta^{n_q^{\textrm{in}}} e^{-ik_{m\omega} r^*} \\
      &\hspace{3em}+ Z_{lm\omega p}^{\textrm{out}} \pb{q} S^{\textrm{out}}_{lm\omega p} (\theta) \Delta^{n_q^{\textrm{out}}} e^{ik_{m\omega} r^*}
    \end{aligned} & r^* \to -\infty
  \end{cases},
\end{equation}
for some constants $n_q^{\textrm{in}/\textrm{out}/\textrm{up}/\textrm{down}}$ that determine the falloffs of $q$, and are (effectively) given in Table~\ref{table:em_asymp}.
The asymptotic angular dependences are given in equations~\eqref{eqn:em_asymp_theta}.

Our final results of this section are the fluxes of $\pb{\pm 1} j^a [\overline{\pb{1} \var \boldsymbol{A}}]$, which are given by

\begin{subequations} \label{eqn:fluxes_pm_1_Q}
  \begin{align}
    \left.\frac{\ud^2 \pb{+1} Q_{ll'm\omega pp'}}{\ud u \ud \Omega}\right|_S &= -\frac{ipp' B_{lm\omega} B_{l'm\omega}}{16} \begin{dcases}
      \Sigma_+ \Im\left[Z_{lm\omega p}^{\textrm{out}} \overline{Z_{l'm\omega p'}^{\textrm{out}}} \pb{\pb{1} \var A_n} S_{lm\omega p}^{\textrm{out}} (\theta) \overline{\pb{\pb{1} \var \bar{\chi}_1} S_{l'm\omega p'}^{\textrm{out}} (\theta)}\right] & S = H^- \\
      \Im\left[Z_{lm\omega p}^{\textrm{up}} \overline{Z_{l'm\omega p'}^{\textrm{up}}} \pb{\pb{1} \var A_n} S_{lm\omega p}^{\textrm{up}} (\theta) \overline{\pb{\pb{1} \var \bar{\chi}_1} S_{l'm\omega p'}^{\textrm{up}} (\theta)}\right] & S = \scri^+ \\
    \end{dcases}, \\
    \left.\frac{\ud^2 \pb{+1} Q_{ll'm\omega pp'}}{\ud v \ud \Omega}\right|_S &= \frac{ip B_{lm\omega} (B_{l'm\omega})^2}{256} \pb{1} \Theta_{l'm\omega} (\theta) \begin{dcases}
      \Im\left[Z_{lm\omega p}^{\textrm{in}} \overline{Z_{l'm\omega p'}^{\textrm{in}}} \pb{\pb{1} \var A_{\bar{m}}} S_{lm\omega p}^{\textrm{in}} (\theta)\right] & S = H^+ \\
      \Im\left[Z_{lm\omega p}^{\textrm{down}} \overline{Z_{l'm\omega p'}^{\textrm{down}}} \pb{\pb{1} \var A_{\bar{m}}} S_{lm\omega p}^{\textrm{down}} (\theta)\right] & S = \scri^-
    \end{dcases}, \\
    \left.\frac{\ud^2 \pb{-1} Q_{ll'm\omega pp'}}{\ud u \ud \Omega}\right|_S &= \frac{ip B_{lm\omega} (B_{l'm\omega})^2}{128} \pb{-1} \Theta_{l'm\omega} (\theta) \begin{dcases}
      \Sigma_+ \Im\left[Z_{lm\omega p}^{\textrm{out}} \overline{Z_{l'm\omega p'}^{\textrm{out}}} \pb{\pb{-1} \var A_m} S_{lm\omega p}^{\textrm{out}} (\theta)/\zeta_+^2\right] & S = H^- \\
      \Im\left[Z_{lm\omega p}^{\textrm{up}} \overline{Z_{l'm\omega p'}^{\textrm{up}}} \pb{\pb{-1} \var A_m} S_{lm\omega p}^{\textrm{up}} (\theta)\right] & S = \scri^+
    \end{dcases}, \\
    \left.\frac{\ud^2 \pb{-1} Q_{ll'm\omega pp'}}{\ud v \ud \Omega}\right|_S &= -\frac{ipp' B_{lm\omega} B_{l'm\omega}}{32} \begin{dcases}
      \Im\left[Z_{lm\omega p}^{\textrm{in}} \overline{Z_{l'm\omega p'}^{\textrm{in}}} \pb{\pb{-1} \var A_l} S_{lm\omega p}^{\textrm{in}} (\theta) \overline{\pb{\pb{1} \var \bar{\chi}_1} S_{l'm\omega p'}^{\textrm{in}} (\theta)}\right] & S = H^+ \\
      \Im\left[Z_{lm\omega p}^{\textrm{down}} \overline{Z_{l'm\omega p'}^{\textrm{down}}} \pb{\pb{-1} \var A_l} S_{lm\omega p}^{\textrm{down}} (\theta) \overline{\pb{\pb{1} \var \bar{\chi}_1} S_{l'm\omega p'}^{\textrm{down}} (\theta)}\right] & S = \scri^-
    \end{dcases},
  \end{align}
\end{subequations}
where $\Sigma^+ \equiv |\zeta_+|^2$, where $\zeta_+ \equiv r_+ + ia \cos \theta$.

\section{Conclusions}

We have found a conserved current for electromagnetic fields whose conserved charge reduces to a sum of positive powers of Carter constants for a stream of photons in the geometric optics limit, and has finite fluxes at infinity.
In these ways, this current generalizes the conserved current for a complex scalar field derived by Carter~\cite{PhysRevD.16.3395}.
In a future paper, we will provide a similar analysis of conserved currents in linearized gravity.
We also plan to explore the interactions between these currents and a charged worldline in order to determine if useful information about the trajectory of the body can be determined from fluxes of these currents.

\section*{Acknowledgements}

The authors acknowledge the support of NSF Grants No. PHY-1404105 and PHY-1707800 to Cornell University.

\appendix

\section{Asymptotic behaviour} \label{appendix:asymptotic}

In this appendix we prove the asymptotic falloff behaviour and angular dependence of the vector potentials and the middle Maxwell scalar $\var \bar{\chi}_1$ that are used in Section~\ref{sec:fluxes} using techniques derived in~\cite{chandrasekhar1983mathematical}.
First, we write these quantities in terms of differential operators acting upon Debye potentials.
Writing out equations~\eqref{eqn:tau_dagger} (for the vector potentials) and~\eqref{eqn:1_bar_chi} (for $\var \bar{\chi}_1$) in Boyer-Lindquist coordinates, we find that

\begin{subequations} \label{eqn:too_tired}
  \begin{gather}
    (\pb{1} \var A)_n = -\frac{1}{2\sqrt{2} \bar{\zeta}} \left(\mathscr{L}_1^+ - \frac{ia \sin \theta}{\zeta}\right) \pb{-1} \psi, \qquad (\pb{1} \var A)_{\bar{m}} = -\frac{1}{2} \left(\mathscr{D}_0 - \frac{1}{\zeta}\right) \pb{-1} \psi, \\
    (\pb{-1} \var A)_l = \frac{\zeta}{2\sqrt{2}} \left(\mathscr{L}_1 - \frac{ia \sin \theta}{\zeta}\right) \pb{1} \psi, \qquad (\pb{-1} \var A)_m = -\frac{\zeta}{4\bar{\zeta}} \left(\mathscr{D}_0^+ - \frac{1}{\zeta}\right) \Delta \pb{1} \psi, \\
    \pb{1} \var \bar{\chi}_1 = -\frac{1}{2\sqrt{2} \bar{\zeta}} \left[\left(\mathscr{L}_1^+ - \frac{ia \sin \theta}{\bar{\zeta}}\right) \mathscr{D}_0 - \frac{1}{\bar{\zeta}} \mathscr{L}_1^+ - \frac{ia \sin \theta}{\zeta} \left(\frac{1}{\zeta} - \frac{2}{\bar{\zeta}}\right)\right] \pb{-1} \psi.
  \end{gather}
\end{subequations}
Since each term in the expansions of $\pb{1} \var \bar{\chi}_0$ and $\bar{\zeta}^2 \pb{1} \var \bar{\chi}_2$ are proportional to the terms in the expansion of $\pb{1} \psi$ [as in equation~\eqref{eqn:debye_def}], their asymptotic behaviour is apparent from the asymptotics in equation~\eqref{eqn:debye_expansion}.

In order to compute the asymptotic behaviour of the vector potentials and $\var \bar{\chi}_1$, we use the asymptotic behaviour of derivatives of the Debye potential $\pb{\pm 1} \psi$.
However, applying the na\"ive approach, which uses the asymptotic expansions given in equation~\eqref{eqn:debye_expansion}, along with

\begin{subequations} \label{eqn:naive_asymptotics}
  \begin{align}
    &\left.\begin{aligned}
        \mathscr{D}_{0(\pm m)(\pm \omega)} f(r) e^{\pm i\omega r^*} &= \frac{\ud f}{\ud r} e^{\pm i\omega r^*} \\
        \mathscr{D}_{0(\pm m)(\pm \omega)} f(r) e^{\mp i\omega r^*} &= \left[\frac{\ud f}{\ud r} \mp 2i\omega f(r)\right] e^{\mp i\omega r^*} \\
      \end{aligned}\right\} r^* \to \infty \\
    &\left.\begin{aligned}
        \mathscr{D}_{0(\pm m)(\pm \omega)} f(r) e^{\pm ik_{m\omega} r^*} &= \frac{\ud f}{\ud r} e^{\pm i\omega r^*} \\
        \mathscr{D}_{0(\pm m)(\pm \omega)} f(r) e^{\mp ik_{m\omega} r^*} &= \left[\frac{\ud f}{\ud r} \mp \frac{4Mr^+}{\Delta} ik_{m\omega} f(r)\right] e^{\mp i\omega r^*} \\
      \end{aligned}\right\} r^* \to -\infty
  \end{align}
\end{subequations}
results in a cancellation in the leading-order behaviour, and so subleading corrections are required.
This issue can be side-stepped using the radial Teukolsky-Starobinsky identities, as follows: first, consider some function $\pb{\pm 1} R_{lm\omega} (r)$ that is a solution to the radial Teukolsky equation~\eqref{eqn:radial_teukolsky} and the radial Teukolsky-Starobinsky identity~\eqref{eqn:radial_tsi_no_p}.
Define

\begin{equation}
  \pb{\pm 1} U_{lm\omega} = \mp 2 i\omega r + \pb{1} \lambda_{lm\omega}, \pb{\pm 1} V_{m\omega} = \mp 2i K_{m\omega},
\end{equation}
such that the radial Teukolsky equation~\eqref{eqn:radial_teukolsky} takes the form

\begin{equation}
  \Delta \mathscr{D}_{0(\mp m)(\mp \omega)}^2 \Delta^{(1 \pm 1)/2} \pb{\pm 1} R_{lm\omega} (r) = \left(\pb{\pm 1} U_{lm\omega} + \pb{\pm 1} V_{m\omega} \mathscr{D}_{0(\mp m)(\mp \omega)}\right) \Delta^{(1 \pm 1)/2} \pb{\pm 1} R_{lm\omega} (r).
\end{equation}
Using equation~\eqref{eqn:radial_tsi_no_p} on the left-hand side, one finds that~\cite{chandrasekhar1983mathematical}:

\begin{equation} \label{eqn:chandrasekhar}
  \mathscr{D}_{0(\mp m)(\mp \omega)} \Delta^{(1 \pm 1)/2} \pb{\pm 1} R_{lm\omega} (r) \equiv \pb{\pm 1} \Xi_{lm\omega} \Delta^{(1 \pm 1)/2} \pb{\pm 1} R_{lm\omega} (r) + \pb{\pm 1} \Pi_{lm\omega} \Delta^{(1 \mp 1)/2} \pb{\mp 1} R_{lm\omega} (r),
\end{equation}
where

\begin{subequations}
  \begin{align}
    \pb{\pm 1} \Xi_{lm\omega} &= -\frac{\pb{\pm 1} U_{lm\omega}}{\pb{\pm 1} V_{m\omega}} = \pm \frac{\pb{1} \lambda_{lm\omega} \mp 2i\omega r}{2iK_{m\omega}} = \begin{dcases}
      \frac{1}{r} \left[1 \pm \frac{i \pb{1} \lambda_{lm\omega}}{2\omega r} + O(1/r^2)\right] & r^* \to \infty \\
      \mp \frac{\pb{s} \lambda_{lm\omega} \mp 2i\omega r^+}{4iMr^+ k_{m\omega}} & r^* \to -\infty
    \end{dcases}, \\
    \pb{\pm 1} \Pi_{lm\omega} &= \frac{2^{\pm 1} B_{lm\omega}}{\pb{\pm} V_{m\omega}} = \mp \frac{B_{lm\omega}}{2^{1 \mp 1} i K_{m\omega}} = \begin{dcases}
      \pm \frac{B_{lm\omega}}{2^{1 \mp 1} i\omega} \frac{1}{r^2} & r^* \to \infty \\
      \pm \frac{B_{lm\omega}}{2^{2 \mp 1} iMr^+ k_{m\omega}} & r^* \to -\infty
    \end{dcases}.
  \end{align}
\end{subequations}
Using equations~\eqref{eqn:naive_asymptotics} and~\eqref{eqn:chandrasekhar}, we furthermore have that

\begin{equation}
  \begin{aligned}
    \pb{1} Z_{lm\omega}^{\textrm{down}} = -\frac{8\omega^2}{B_{lm\omega}} \pb{-1} Z_{lm\omega}^{\textrm{down}}, &\qquad \pb{-1} Z_{lm\omega}^{\textrm{up}} = -\frac{2\omega^2}{B_{lm\omega}} \pb{1} Z_{lm\omega}^{\textrm{up}}, \\
    \pb{1} Z_{lm\omega}^{\textrm{in}} = -\frac{32 M^2 r_+^2 k_{m\omega}^2 \pb{-1} \kappa_{m\omega}}{B_{lm\omega}} \pb{-1} Z_{lm\omega}^{\textrm{in}}, &\qquad \pb{-1} Z_{lm\omega}^{\textrm{out}} = -\frac{8 M^2 r_+^2 k_{m\omega}^2 \pb{1} \kappa_{m\omega}}{B_{lm\omega}} \pb{1} Z_{lm\omega}^{\textrm{out}},
  \end{aligned}
\end{equation}
where

\begin{equation}
  \pb{s} \kappa_{m\omega} = 1 - \frac{is(r_+ - M)}{2Mr_+ k_{m\omega}}.
\end{equation}

\begin{table}
  \centering
  \begin{tabular}{|l|l l|l l|} \hline
    & \multicolumn{2}{c|}{Ingoing} & \multicolumn{2}{c|}{Outgoing} \\
    & $r^* \to -\infty$ & $r^* \to \infty$ & $r^* \to -\infty$ & $r^* \to \infty$ \\
    & $e^{-ik_{m\omega} r^*} \times$ & $e^{-i\omega r^*} \times$ & $e^{ik_{m\omega} r^*} \times$ & $e^{i\omega r^*} \times$ \\\hline
    $\pb{1} \var A_n$ & $\Delta$ & $1/r^2$ & $1$ & $1$ \\
    $\pb{1} \var A_{\bar{m}}$ & $1$ & $1/r$ & $1$ & $1/r$ \\
    $\pb{-1} \var A_l$ & $1/\Delta$ & $1$ & $1$ & $1/r^2$ \\
    $\pb{-1} \var A_m$ & $1$ & $1/r$ & $1$ & $1/r$ \\
    $\pb{1} \var \bar{\chi}_0$ & $1/\Delta$ & $1/r$ & $1$ & $1/r^3$ \\
    $\pb{1} \var \bar{\chi}_1$ & $1$ & $1/r^2$ & $1$ & $1/r^2$ \\
    $\pb{1} \var \bar{\chi}_2$ & $\Delta$ & $1/r^3$ & $1$ & $1/r$ \\\hline
  \end{tabular}
  \caption{\label{table:em_asymp} Asymptotic behaviour of the solutions for electromagnetism.}
\end{table}

Combining these asymptotic formulas with equations~\eqref{eqn:too_tired} yields the asymptotic falloffs given in table~\ref{table:em_asymp}, as well as the following angular factors:

\begin{subequations} \label{eqn:em_asymp_theta}
  \begin{align}
    \pb{\pb{1} \var A_n} S^{\textrm{in}}_{lm\omega p} (\theta) &= \pb{\pb{1} \var A_n} S^{\textrm{out}}_{lm\omega p} (\theta) = -\frac{1}{2\sqrt{2}} \mathscr{L}^+_{1m\omega} \pb{-1} \Theta_{lm\omega} (\theta), \\
    \pb{\pb{1} \var A_n} S^{\textrm{down}}_{lm\omega p} (\theta) &= \pb{\pb{1} \var A_n} S^{\textrm{up}}_{lm\omega p} (\theta) = -\frac{1}{2\sqrt{2} \zeta_+} \left(\mathscr{L}^+_{1m\omega} - \frac{ia \sin \theta}{\zeta^+}\right) \pb{-1} \Theta_{lm\omega} (\theta), \\
    \pb{\pb{1} \var A_{\bar{m}}} S^{\textrm{in}}_{lm\omega p} (\theta) &= 2Mr_+ ik_{m\omega} \pb{-1} \kappa_{m\omega} \pb{-1} \Theta_{lm\omega} (\theta), \\
    \pb{\pb{1} \var A_{\bar{m}}} S^{\textrm{out}}_{lm\omega p} (\theta) &= \frac{1}{2} \left(\frac{\pb{1} \lambda_{lm\omega} + 2i\omega r_+}{4iMr_+ k_{m\omega}} + \frac{1}{\zeta_+}\right) \pb{-1} \Theta_{lm\omega} (\theta), \\
    \pb{\pb{1} \var A_{\bar{m}}} S^{\textrm{down}}_{lm\omega p} (\theta) &= i\omega \pb{-1} \Theta_{lm\omega} (\theta), \\
    \pb{\pb{1} \var A_{\bar{m}}} S^{\textrm{up}}_{lm\omega p} (\theta) &= \frac{i}{2} \left(\frac{\pb{1} \lambda_{lm\omega}}{2\omega} + a \cos \theta\right) \pb{-1} \Theta_{lm\omega} (\theta), \\
    \pb{\pb{-1} \var A_l} S^{\textrm{in}}_{lm\omega p} (\theta) &= \pb{\pb{-1} \var A_l} S^{\textrm{out}}_{lm\omega p} (\theta) = \frac{\zeta_+}{2\sqrt{2}} \left(\mathscr{L}_{1m\omega} - \frac{ia \sin \theta}{\zeta^+}\right) \pb{1} \Theta_{lm\omega}, \\
    \pb{\pb{-1} \var A_l} S^{\textrm{down}}_{lm\omega p} (\theta) &= \pb{\pb{-1} \var A_l} S^{\textrm{up}}_{lm\omega p} (\theta) = \frac{1}{2\sqrt{2}} \mathscr{L}_{1m\omega} \pb{1} \Theta_{lm\omega} (\theta), \\
    \pb{\pb{-1} \var A_m} S^{\textrm{in}}_{lm\omega p} (\theta) &= -\frac{\zeta_+}{4\bar{\zeta}_+} \left(\frac{\pb{1} \lambda_{lm\omega} - 2i\omega r_+}{4iMr_+ k_{m\omega}} - \frac{1}{\zeta_+}\right) \pb{1} \Theta_{lm\omega} (\theta), \\
    \pb{\pb{-1} \var A_m} S^{\textrm{out}}_{lm\omega p} (\theta) &= -\frac{\zeta_+}{\bar{\zeta}_+} Mr_+ ik_{m\omega} \pb{1} \kappa_{m\omega} \pb{1} \Theta_{lm\omega} (\theta), \\
    \pb{\pb{-1} \var A_m} S^{\textrm{down}}_{lm\omega p} (\theta) &= -\frac{i}{4} \left(\frac{\pb{1} \lambda_{lm\omega}}{2\omega} - a\cos \theta\right) \pb{1} \Theta_{lm\omega} (\theta), \\
    \pb{\pb{-1} \var A_m} S^{\textrm{up}}_{lm\omega p} (\theta) &= -\frac{i}{2} \omega \pb{1} \Theta_{lm\omega} (\theta), \\
    \pb{\pb{1} \var \bar{\chi}_1} S^{\textrm{in}}_{lm\omega p} (\theta) &= \frac{\sqrt{2}}{\bar{\zeta}_+} Mr_+ ik_{m\omega} \left(\mathscr{L}_{1m\omega}^+ - \frac{ia \sin \theta}{\bar{\zeta}_+}\right) \pb{-1} \Theta_{lm\omega} (\theta), \\
    \pb{\pb{1} \var \bar{\chi}_1} S^{\textrm{out}}_{lm\omega p} (\theta) &= -\frac{1}{2\bar{\zeta}_+ \sqrt{2}} \Bigg\{\left(\frac{\pb{1} \lambda_{lm\omega} + 2i\omega r_+}{4iMr_+ k_{m\omega}} - \frac{1}{\bar{\zeta}_+}\right) \nonumber \\
                                                               &\hspace{6em}- ia \sin \theta\left[\frac{1}{\bar{\zeta}_+} \frac{\pb{1} \lambda_{lm\omega} + 2i\omega r_+}{4iMr_+ k_{m\omega}} + \frac{1}{\zeta_+} \left(\frac{1}{\zeta_+} - \frac{2}{\bar{\zeta}_+}\right)\right]\Bigg\} \pb{-1} \Theta_{lm\omega} (\theta), \\
    \pb{\pb{1} \var \bar{\chi}_1} S^{\textrm{down}}_{lm\omega p} (\theta) &= \frac{i\omega}{\sqrt{2}} \mathscr{L}_{1m\omega}^+ \pb{-1} \Theta_{lm\omega} (\theta) , \\
    \pb{\pb{1} \var \bar{\chi}_1} S^{\textrm{up}}_{lm\omega p} (\theta) &= \frac{i}{2\sqrt{2}} \left(\frac{\pb{1} \lambda_{lm\omega}}{2\omega} + ia \cos \theta\right) \mathscr{L}_{1m\omega}^+ \pb{-1} \Theta_{lm\omega} (\theta).
  \end{align}
\end{subequations}

\bibliography{adwg}

\end{document}